%% file: QReach.tex
\documentclass[10pt,conference,letterpaper]{IEEEtran}
\IEEEoverridecommandlockouts
\usepackage{cite}
\usepackage{amsmath,amssymb,amsfonts}
\usepackage[font=small]{caption}
\usepackage{algorithmic}
\usepackage{stfloats}
\usepackage{blindtext}
\usepackage{subfig}
\usepackage{resizegather}
\usepackage{graphicx}
\usepackage{textcomp}
\usepackage{xcolor}
\usepackage{tabu}
\usepackage{url}
\usepackage{lipsum}
\usepackage{multicol}
\usepackage{mathtools}
\usepackage{cuted}
\usepackage[normalem]{ulem}
\usepackage{amsthm}
\usepackage[utf8]{inputenc}
\usepackage{fancyhdr,graphicx}
\usepackage[ruled,vlined]{algorithm2e}
\usepackage{array}
\usepackage{multirow}
\newcommand{\name}{\text{Q-REACH}}
\def\BibTeX{{\rm B\kern-.05em{\sc i\kern-.025em b}\kern-.08em
    T\kern-.1667em\lower.7ex\hbox{E}\kern-.125emX}}

\newtheorem{observation}{Observation}

\usepackage{tikz} % for drawing the digital logic gates
\usetikzlibrary{arrows,shapes.gates.logic.US,shapes.gates.logic.IEC,calc}
\newtheorem{assumption}{Assumption}

\begin{document}

% \title{Quantum Teleportation Error-Mitigation with  Bit-Phase-Flip}
% \title{ Q-Rector: \underline{Q}uantum Error Correction  in Quantum \underline{Re}peater  \underline{C}aching Ne\underline{t}w\underline{or}k}

\title{
%\underline{Q}uantum \underline{Re}peater \underline{C}aching Network Pro\underline{t}ocol Design for Quantum Err\underline{o}r Co\underline{r}rection 
Q-REACH: \underline{Q}uantum information \underline{R}epetition, \underline{E}rror \underline{A}nalysis and \underline{C}orrection using Cac\underline{h}ing Network}

\author{
\IEEEauthorblockN{Karl C. Linne (Kai Li)\textsuperscript{1}, Yuanyuan Li\textsuperscript{2}, Debashri Roy\textsuperscript{3}, Kaushik Chowdhury\textsuperscript{4}}

\IEEEauthorblockA{\textsuperscript{1}Pritzker School of Molecular Engineering , The University of Chicago, Chicago, IL, USA}
\IEEEauthorblockA{\textsuperscript{2}Department of Electrical and Computer Engineering, Northeastern University, Boston, MA, USA}
\IEEEauthorblockA{\textsuperscript{3}Department of Computer Science and Engineering, The University of Texas at Arlington, Arlington, TX, USA}
\IEEEauthorblockA{\textsuperscript{4}Chandra Family Department of Electrical and Computer Engineering, The University of Texas at Austin, Austin, TX, USA}
}
\maketitle

\begin{abstract}
Quantum repeaters incorporating quantum memory play a pivotal role in mitigating loss in transmitted quantum information (photons) due to link attenuation over a long-distance quantum communication network. However, limited availability of available storage in such quantum repeaters and the impact on the time spent within the memory unit presents a trade-off between \textcolor{black}{quantum information} fidelity (a metric that quantifies the degree of similarity between a pair of quantum states) and qubit transmission rate. Thus, effective management of storage time for qubits becomes a key consideration in multi-hop quantum networks. To address these challenges, we propose $\name$, which leverages queuing theory in caching networks to tune qubit transmission rate while considering fidelity as the cost metric. Our contributions in this work include (i) utilizing a method of repetition that encodes and broadcasts multiple qubits through different quantum paths, (ii) analytically estimating the time spent by these emitted qubits as a function of the number of paths and repeaters, as well as memory units within a repeater, and (iii) formulating optimization problem that leverages this analysis to correct the transmitted logic qubit and select the optimum repetition rate at the transmitter.

\end{abstract}


\input{section/introduction.tex}
\input{section/background.tex}

\input{section/theory.tex}

\input{section/result.tex}

% \input{section/application.tex}
\input{section/conclusion.tex}

% \input{section/acknowledgements.tex}

% \bibliographystyle{IEEEtran}
% \bibliography{reference.bib}

\input{QReach_reference.bbl}

\end{document}

%% file: section/introduction.tex
\section{Introduction}
\vspace*{-3pt}
% \IEEEPARstart{O}{bject} detection and tracking based on both far and near electromagnetic fields have been studied extensively over the past several years.
The successful implementation of quantum applications, such as quantum key distribution QKD~\cite{qkd}, teleportation~\cite{teleportation}, entanglement distribution\cite{zhao2022e2e, zeng2022multi}, distributed quantum computation~\cite{yuanyuan, qiao_1,Li2023},  sensing~\cite{dist_sensing}, and clock synchronization~\cite{quantum_clock}, hinges upon the establishment of a fault-tolerant quantum communication system that can operate at large separation distances. Quantum memory-enabled quantum repeaters have emerged as a viable solution to combat the fragile nature of qubits over quantum links\cite{chaudhary2023learning} and their exponential attenuation with transmission length. 

While such repeaters maintain transmitted qubit fidelity
% (a metric that quantifies the degree of similarity between a pair of quantum states) 
by preventing the so-called quantum \textit{decoherence} (loss of quantum coherence due to the interaction with the environment)\cite{li2023bip} in the state of the qubit during single-hop transmissions, they in turn raise two key challenges: First, the available quantum memory is limited at a given repeater, and this requires careful determination of how many qubits should be injected in the network at a given time. 
Second, the impact of \textit{decoherence} continues during the time a given qubit is buffered in the quantum memory. The above two issues do not permit simply increasing the number of qubits to enhance error correction at the end-destination. This is because increasing network traffic with limited memory in the repeater increases queuing delay, which in turn degrades the quality of each individual qubit. To address these challenges, we propose $\name$, which analyzes the end-to-end performance of such a quantum communication system using tools from queuing theory in a repeater network, analogous to a caching network. An accurate estimate of the time spent by a qubit in the overall network then allows $\name$ to perform error correction at the end-destination for  resilient end-to-end transmission.

%The fragile nature of quantum bits (qubits) over quantum link,  exponential attenuation with transmission length, motivates the widely application of quantum memory assisted quantum repeater to establish quantum data transmission link at-a-distance, and different qubit error correction methods. However, the limited storage time of quantum memory for effectively maintaining the fidelity of qubit imposes critical challenges on the performance of quantum communication system, end-to-end quantum data transmission rate and fidelity in particular. Additionally, the time-sensitive nature of quantum memory is challenging the optimal overhead on the corresponding quantum error correction method over the complex quantum communication network with multiple quantum repeaters. Therefore, developing an effective caching network for queuing the qubit transmission through precise  waiting control of quantum memory and designing a more powerful quantum error correction method for correcting multiple errors at one time are of utmost importance and a keystone towards fault-tolerant, large-scale quantum communication network.  

\begin{figure}[t!]
\centering
\includegraphics[width=1.0\linewidth]{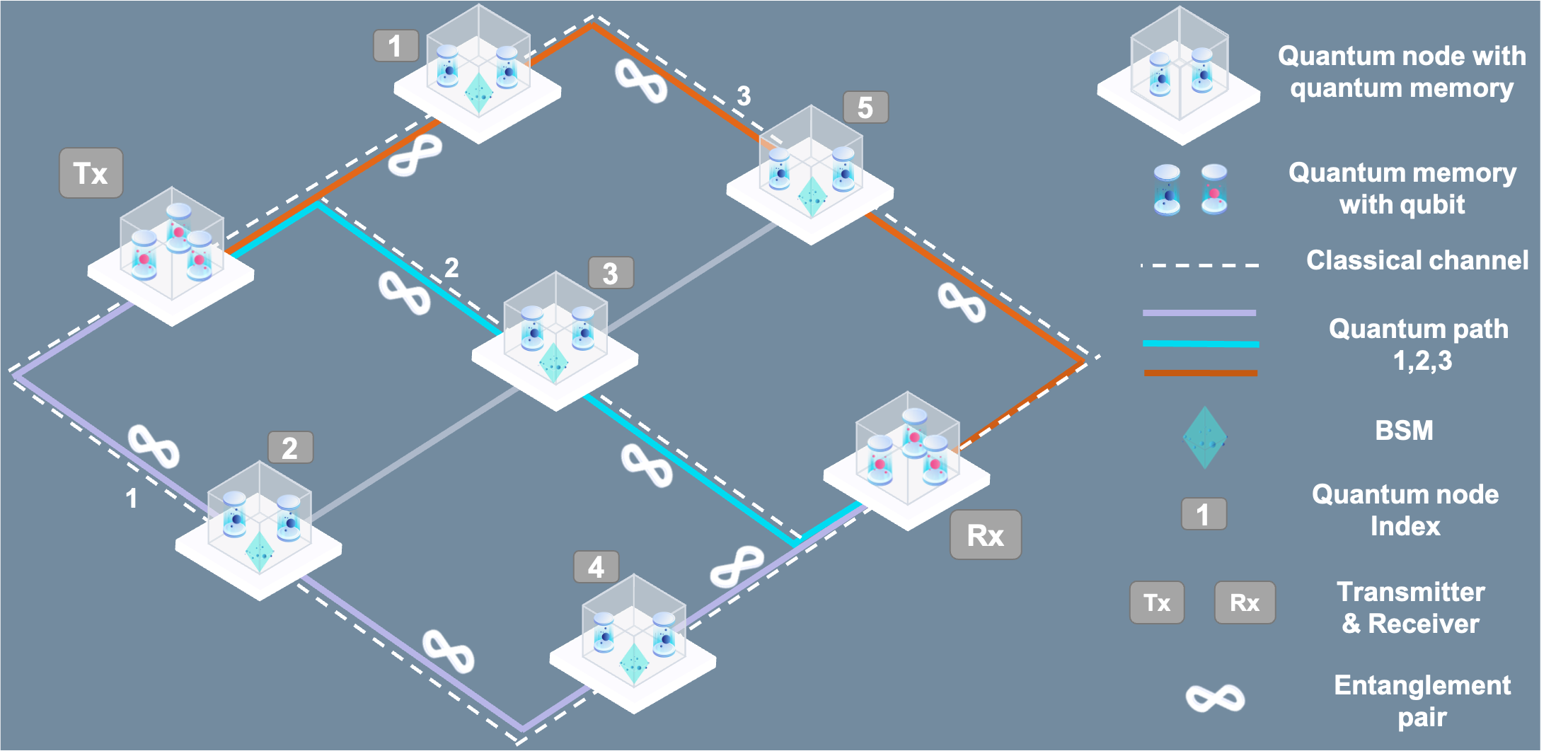}
\caption{Overview of the proposed quantum caching network. The encoded logical qubit transfer physical qubits through different paths $1, 2, 3$, each path includes multiple quantum repeaters and each quantum repeater is equipped with multiple quantum memory units.}
\vspace*{-15pt}
\label{fig:1st}
\end{figure}

\noindent$\bullet$\textbf{Limitations in the current state-of-the-art:} Quantum memory facilitates precise entanglement swapping operations in quantum repeaters by synchronizing two qubits. Since the entanglement swapping only occurs at the head-of-the-line qubit, the subsequently arriving qubits must be accommodated in the limited memory of the repeater, and thus, they incur additional waiting time. %, leveraging the additional waiting time provided by storing qubits in the quantum memory. 
Prior work has explored the impact of multiple such entanglement swapping operations and the calculation of the cumulative time for arrival as a single qubit traverses the end-to-end path~\cite{waiting_1, waiting_2}. This time has also been leveraged as a metric within routing algorithms to maximize the overall qubit transmission rate~\cite{waiting_qiao,li_waiting}. Other explorations in this area of work include use of Markov chains to estimate the waiting time distribution over each path and to minimize the queuing length with dynamic programming~\cite{dai2020quantum}.

Despite making significant strides, prior work does not jointly study the design of a repeater in terms of the impact of the number of memory units and the end-to-end queuing delay. Furthermore, a complete error correction framework has not yet been developed that uses this timing analysis and the ensuing degradation in the metric of qubit fidelity to decide how many qubits to inject at the source and how the error correction must be undertaken at the destination. Intuitively, while repeatedly transmitting more qubits can help with error recovery at the destination, they also increase the average waiting time in the repeater queues, which in turn adversely impacts decoding for any single qubit considered in isolation. $\name$ attempts to balance these opposing outcomes.

%Though the aforementioned waiting time related algorithm and the protocol are promising, they have several underlying shortcomings: 1), the existing algorithm or protocol for effective entanglement swapping focus on the maximization of end-to-end qubit transmission throughput without the consideration of queuing performance of each transmitted qubit; 2)  current quantum queuing protocols place significant emphasis on qubit transmission traffic management and congestion control through effective timing schedules, however allocate less attention to the analysis of corresponding qubit loss, 3) both the approaches for the waiting time processing in these two scenarios are limited in exploring and characterizing the performance in the more complicate and practical quantum repeater network, and 4)in pursuit of establishing a robust quantum network, the analysis of quantum error correction is regrettably not addressed within the context of the waiting time analysis     

\noindent$\bullet$\textbf{Challenges and need for new tools:} In repetition-based quantum error correction, the encoding of a single state of the logical qubit (say, polarization of a photon) with multiple physical qubits increases the probability of recovering the correcting encoding state. % is followed by the critical task of synchronizing the states of these physical qubits, which plays a significant role in reducing errors while minimizing the required waiting time. However,  the time-sensitive nature of quantum memory and the additional queuing time during qubit transmission pose significant challenges. 
As this is analogous to the effect of packet caching in classical networks, we adopt tools from queuing theory that have been successfully applied to analyze caching networks \cite{li2020universally}. These tools influence the design of classical networks so that data packets may be queued effectively in node buffers, minimize the waiting time for servers, and thereby enhance data transmission rates. However, the application of classical caching network theory for quantum error correction is non-trivial for the following challenges: (C1) \textit{Fewer samples:} The limited allowed storage time and capacity of the quantum memory in each quantum repeater is far smaller than that in a classical data network (i.e., single-digit qubits can be stored versus hundreds or thousands of classical data packets). This requires careful computation of waiting time distributions since there are few samples to derive this metric. 
(C2) \textit{End-to-end latency:} The impact of waiting-time within the quantum memory and the fragile nature of the qubit during the transmission, coupled with the `no cloning' rule of quantum communication, makes the computation of end-to-end latency difficult, requiring incorporation of deep quantum mechanical equations in the analysis. (C3) \textit{Joint routing and  error correction:} Quantum error correction-based on the repetition code requires encoding quantum data with different number of qubits and then transferring these qubits through different paths to the receiver. The constraints of impact of qubit fidelity on both propagation distance and waiting time within the quantum memories within a given path make the co-design of both the caching network and quantum error correction difficult due to the large state-space of possibilities.

\noindent$\bullet$\textbf{Proposed $\name$ framework and contributions:}  In this work, we propose a queuing theory-based approach that analyzes the end-to-end time a physical qubit remains in the network by leveraging formulations from caching networks, which is then used for quantum error correction. As shown in Fig. \ref{fig:1st}, we use repetition coding at the transmitter (Tx) and physical qubits are transmitted through different paths, such as paths 1, 2, 3, to the receiver. Each path has different number of repeaters, and each repeater has different number of quantum memory units for  storing qubits. While these assumptions make the formulations general, they also complicate the analysis of the end-to-end waiting time and subsequent error correction.  

%We use the $M/D/1/\mathcal{I}$ queuing model for the caching network protocol design, with deterministic serving rate ($D$), Poisson distributed arrival rate ($M$)\cite{vardoyan2019stochastic} for each quantum memory,  1 server for each path,  and $\mathcal{I}$ feasible quantum memories. %The corresponding caching network protocol achieves the following two objectives: 1)  it utilizes the waiting time and number of quantum memories as the new cost metric and variables in each repeaters of each path, 2) creating a benchmark for system performance evaluation with error analysis and correction. 

\noindent$\bullet$\textbf{Summary of contributions:}
The main contributions of the $\name$ are as follows:
\begin{enumerate}
%    \item We provide the physical operation principle of quantum repeater with quantum memory and the analytical correlation between waiting time and the fidelity variation.    
\item We overcome (C1) by including the storage time in each quantum memory unit as one of the cost metrics. This allows us to make a logical connection between the number of quantum memory units within a repeater and the queuing length for each repeater in the path associated with the $M/D/1/\mathcal{I}$ model.  

\item We provide complete theoretical analysis of end-to-end fidelity of received qubits and the encoded logical qubit rate variations as a function of the number of memory units, number of repeaters and the accuracy of the decoded logical qubits to address the challenge (C2).
    
\item  We utilize repetition approach as quantum error correction method of this work. We formulate an optimization problem that selects the logical qubit rate maximization with the quantum error correction performance. This new formulated optimization problem is solved by leveraging analytical predictions using the caching network, which addresses challenge (C3). 
    % \item We envision the utilization of quantum transducer, and provide quantum circuit design to detect and correct multiple qubit errors as the last contribution.
\end{enumerate}

\label{sec:intro}

%% file: section/background.tex
\section{Background }
\vspace*{-3pt}

In this section, we give a foundational background that includes 1) the operating principle of quantum memory and repeater and 2) the quantum error correction method with repetition method. 
% $\mathrm{p}$

% Quantum repeater has been widely studied\cite{}, and is the fundamental component to conduct entangle swapping operation to establish entanglement at long distance. Fig. \ref{fig:repeater} shows the operation mechanism of creating new entanglement pair (q1-q4) with two entanglement (q1-q2, q2-q4) . It is important to emphasise that quantum memory are used as temporally storage to adjustment the start time of BSM due to the different arrival time of each qubit. Similar to the purpose of BSM, quantum memory are used to storage the qubit in the queue to avoid the qubit transmission traffic, which is the key problem we are optimizing in this work by designing caching network.      

\begin{figure}[h!]
\centering
\includegraphics[width=0.70\linewidth]{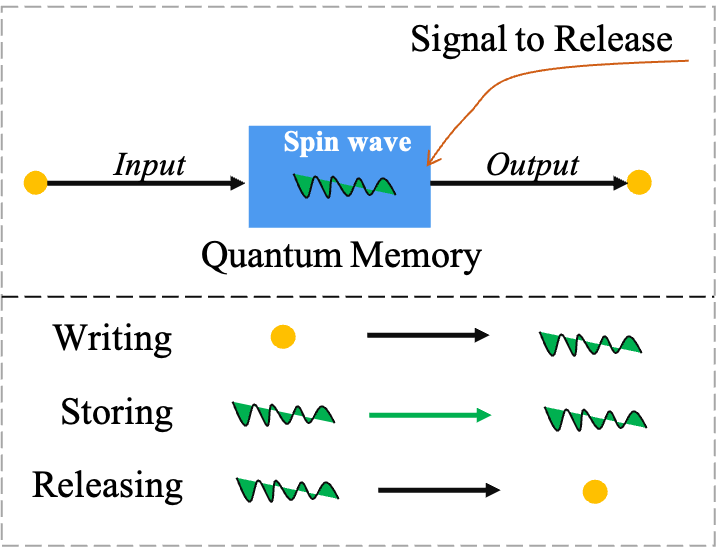}
\caption{Operation of a quantum memory with three steps of writing, storing and releasing. Writing represents converting qubit to spin wave in quantum memory, storing represents the ultra low speed propagation in quantum memory, and the releasing represents converting spin wave to qubit. }
\vspace*{-7pt}
\label{fig:memory}
\end{figure}

\begin{figure}[h!]
\centering
\includegraphics[width=0.9\linewidth]{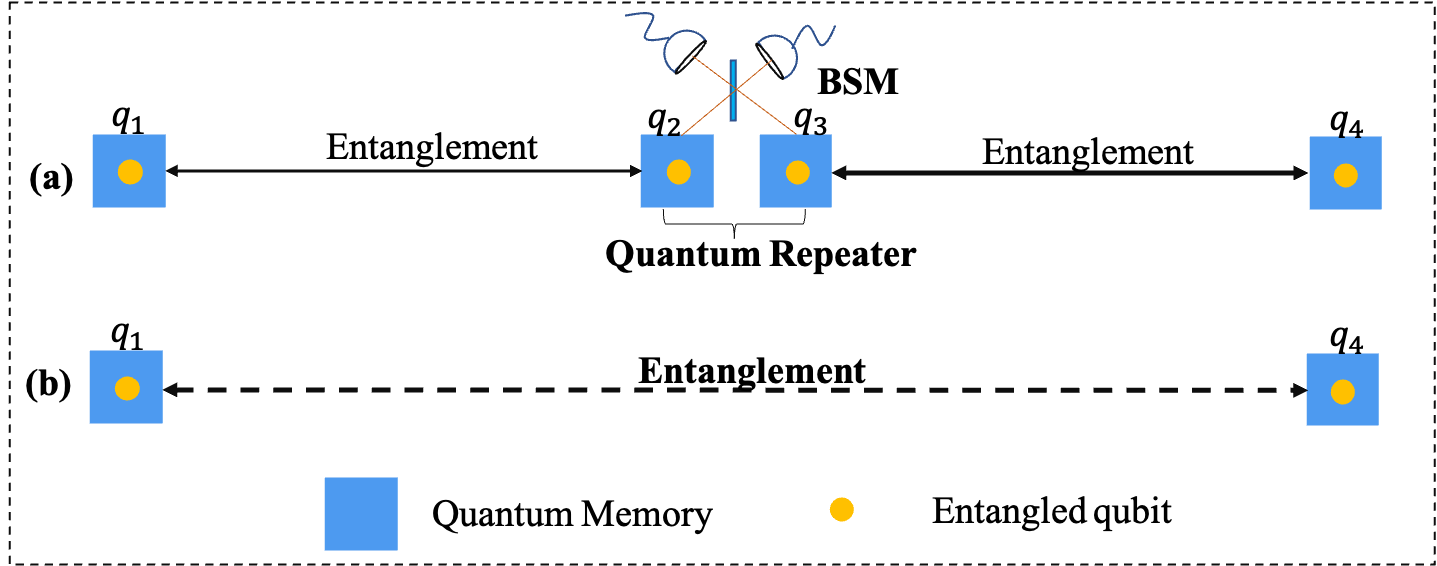}
\caption{Demonstration of a quantum repeater operation with quantum memory. At least four quantum memory units are required for one quantum repeater to establish end-to-end entanglement.}
\vspace*{-10pt}
\label{fig:repeater}
\end{figure}

\noindent$\bullet$\textbf{Operating principle of quantum memory and repeater:}
Fig. \ref{fig:memory} shows how a qubit, represented by a photon, is stored in the quantum memory involving three steps of writing, storing, and releasing~\cite{panigrahy2022capacity}. In the first step, the qubit is transformed into a spin wave within the quantum memory material, such as Rubidium~\cite{gera2020mhz}. The spin wave, characterized by an ultra-low propagation speed, facilitates the storing process. The second step involves the retrieval of the stored information, achieved through a reverse operation that converts the spin wave back into the qubit. In the ideal case, if these operations are executed perfectly, the output qubit is identical to the input qubit. 
%Over time, the fidelity of the output degrades, ultimately as quantum memory experiences losses. 
A comprehensive review of the operation of a quantum memory can be found in \cite{lei2022electromagnetically}. Importantly, within the temporal storage capability of quantum memory two qubits undergo entanglement swapping~\cite{pouryousef2022quantum, farahbakhsh2022opportunistic}. This process involves performing a Bell State Measurement (BSM) of the two initially separated entangled pairs ($q_{2}$ and $q_{3}$), followed by the establishment of new entangled pair between the remaining two individual entangled pairs ($q_{1}$ and $q_{4}$),  as depicted in Fig. \ref{fig:repeater}. During these processes, a total of four quantum memory units are utilized. Two of these memories are used to store the qubits ($q_{2}$ and $q_{3}$) and facilitate their synchronization for the entanglement swapping operation. The other two memory units are utilized to store the qubits ($q_{1}$ and $q_{4}$) for future usage in subsequent steps.

% \noindent$\bullet$\textbf{Queue theory in caching network protocol (\textcolor{red}{Yuanyuan}):}

\noindent$\bullet$\textbf{Quantum error correction method with repetition :} %In the current state of the art, there have been various studies on current multiple errors in a quantum network. 
% Theoretically, the quantum error correction method employs the repetition technique to encode a single logic qubit with multiple physical qubits, effectively reducing the probability of errors in the logic qubit. 
% In 1995, Peter Shor\cite{ shor1995scheme} proposed the pioneering quantum error correction code 
% employing a 9-qubit correction code,
% which draws significant inspiration from the 3-qubit repetition coding method. 
The quantum error correction method is initially developed for present-day noisy intermediate-scale quantum (NISQ) processors by encoding a single logical qubit with multiple physical qubits, with pioneering contributions from 
Peter Shor~\cite{ shor1995scheme}. His work 
 is inspired by the 3-qubit repetition coding method, which encode one logical qubit with 3 physical qubits. 
Most recently, error correction in logical qubits have been demonstrated via the so called surface coding method \cite{gottesman1997stabilizer}, which is an improvement over the original  repetition coding method.
% , which is a topological stabilizer code \cite{gottesman1997stabilizer}, widely considered the best-performing approach for error detection and correction . 
So far, repetition codes have been experimentally demonstrated in quantum computers; however, their practical implementation at a large scale in quantum communication networks still presents significant challenges. The work \cite{luo2021quantum} demonstrated the teleportation of one logical qubit using a polarization-entangled four-photon state. However, the difficulties in generating multiple-party entanglement impacts the system's performance, particularly the quantum data transmission rate. $\name$ addresses this gap in the design of a practically feasible protocol for quantum error correction in large-scale and complex quantum networks.

\tikzstyle{branch}=[fill,shape=circle,minimum size=3pt,inner sep=0pt]

\label{sec:related}
\vspace{-1mm}

%% file: section/theory.tex
\section{Theoretical Analysis and Problem Formulation}
\label{sec:systemdesign}
\vspace*{-3pt}

In this section, we outline the theoretical analysis and problem formulation that is solved by $\name$. First, we present the architecture of the repeater-based quantum network. 
Then, we introduce the error model used to characterize the quantum link (fiber) attenuation and quantum memory loss experienced by qubits during their transmission. Third, we provide a theoretical analysis regarding the edge cost and path cost, considering the time overhead and length overhead as key factors. Fourth, we illustrate the caching network model and formulate the optimization problem of setting the optimum transmission rate using queuing theory in \name.
% At the end, we provide quantum circuit design analysis for multiple quantum error detection and correction.
% 

\begin{figure}[h!]
\centering
\includegraphics[width=1.0\linewidth]{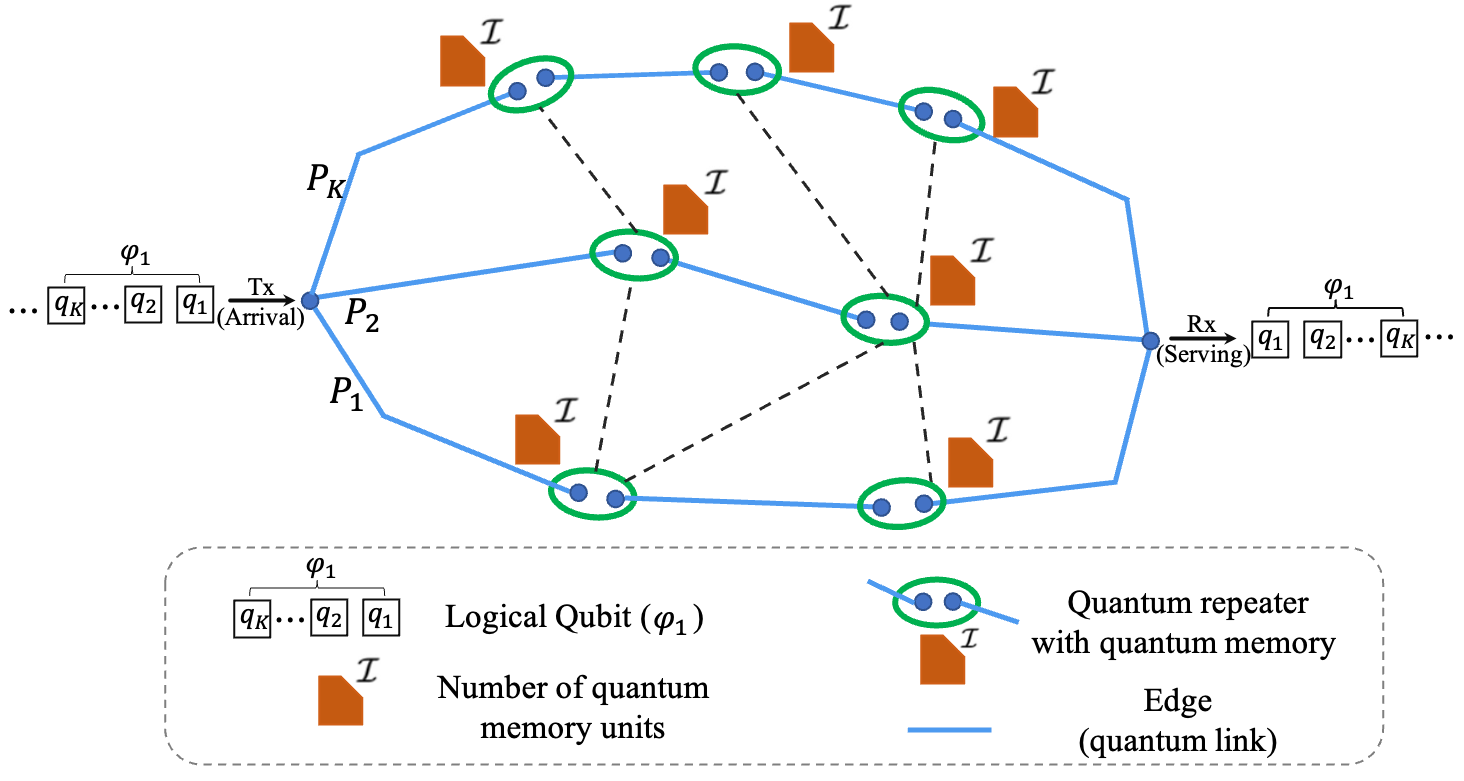}
\caption{A sample network architecture of a quantum network with repeaters, involving $K = 3$ paths, and different number of repeaters and quantum memory units.}
\vspace*{-15pt}
\label{fig:archi}
\end{figure}
\subsection{Network Architecture}
\label{subsec:network_architecture}
The Fig. \ref{fig:archi} shows the architecture of the proposed quantum caching network. Since we use repetition of qubits, or so called repetition coding, a single logical qubit $\varphi_{1}$ is encoded to multiple physical qubits ($q_{1}, q_{2}, ..., q_{K}$) at the transmitter (Tx). The information of these physical qubits is relayed over $K$ paths, denoted as $P = {p_{1}, p_{2}, \cdots, p_{K}}$, to the receiver (Rx). 
Each path consists of $(\mathcal{M}-1)$ quantum repeaters, represented by a set $N$ with cardinality $|N| = (\mathcal{M}-1)$. Along each path, these $\mathcal{M}-1$ quantum repeaters generate a total of $\mathcal{M}$ quantum links, defined as the edge set $E$. Additionally, at each quantum repeater, the set of quantum memory unit is denoted as $C$, where $|C| = \mathcal{I}$. The entanglement swapping operations are performed at each quantum repeater to establish long-distance entanglement between the transmitter (Tx) and the receiver (Rx). This entanglement is utilized to deliver encoded logical qubits from Tx to Rx, which follows the Poisson process for entangled pair generation
\cite{vardoyan2019stochastic}.  As a result, the success of the logical qubit transmission is a function of the entanglement swapping  that occurs between the transmitter (Tx) and receiver (Rx) at each intermediate edge. Throughout the rest of this paper, we refer to the logical qubit as the fundamental metric for evaluating the performance of the entire system. Consequently, the required number of repeaters, edges, and quantum memory units for each path, as well as the total number of paths in the system, can be mathematically represented as a graph: $G=\{P, N, E, C \}$, where $P=\{p_{1}, p_{2},...,p_{K}\}$, $N=\{n_{1}, n_{2},...,n_{\mathcal{M}-1}\}$, $E=\{e_{1}, e_{2},...,e_{\mathcal{M}}\}$, where $C=\{c_{1}, c_{2},...,c_{\mathcal{I}}\}$,  $K$, $(\mathcal{M}-1)$ and $\mathcal{I}$ are the number of paths, repeaters and memory units, respectively. Additionally, we make the following assumptions: 

\begin{assumption}
\vspace*{-5pt}
Each quantum repeater has a varying number of quantum memory units, and each quantum memory unit is capable of storing one qubit at a time.
% \vspace*{-5pt}
\end{assumption}

\begin{assumption}
\vspace*{-5pt}
Quantum repeaters can be interconnected by multiple quantum links, represented by either blue solid lines or black dashed lines in Fig. \ref{fig:archi}. Each quantum link can access different memory units inside the quantum repeater.% at different times.
% \vspace*{-5pt}
\end{assumption}

\begin{assumption} 
\vspace*{-5pt}
The quantum receiver is equipped with multiple detectors, enabling the simultaneous detection of the encoded $K$ physical qubits. % from the $K$ paths.
\vspace*{-5pt}
\end{assumption}

% \begin{assumption}
% For the purposes of envisioning quantum error correction, we assume that the receiver is equipped with an ideal quantum transducer and possesses the capability of quantum computing and quantum circuit designing. 
% \end{assumption}

\subsection{Error Model}

The fragile nature of qubits leads to a degradation of qubit fidelity during transmission. This degradation is primarily caused by two main factors: the attenuation of fiber-based quantum links and the loss of quantum information due to storage time in each quantum memory unit.  
% This results in their degradation. 
Statistically, this degradation is modeled using an error operator $\varepsilon(\wp)$ through different weighted Pauli operators. Following \cite{request_schedule}, the error operator $\varepsilon(\wp)$ for the given state of qubit $|\psi \rangle$ can be  expressed as: 
\begin{equation}
    \varepsilon(\wp) = \sum_{v}\wp_{v}O_{v}\rho O_{v}^{\dagger},
\label{equ:erroioerator}
\end{equation}
where $O_{v}$ represents the $v^{th}$ operator drawn from the options of $X$, $Y$, $Z$, or $I$ Pauli operators, $\rho$ is the density matrix of the given qubit state $|\psi \rangle$, calculated as $\rho = |\psi \rangle \langle \psi|$, and $p_v$ denotes the weight of corresponding operators. These weights are calculated by the probability ($\wp$) of degradation of qubit fidelity due to 1) attenuation in the fiber $\left(\wp_{fiber}\right)$ with length and 2) the storage time in the quantum memory $\left(\wp_{memory}\right)$. Specifically, substituting $\wp_{fiber}$ and $\wp_{memory}$ into Eq.~\eqref{equ:erroioerator}, we get:

\begin{equation}
\begin{split}
\varepsilon(\wp_{fiber})& = (1-0.75\wp_{fiber}) I \rho I^{\dagger} + 0.25\wp_{fiber} X \rho X^{\dagger} \\&+0.25\wp_{fiber} Y \rho Y^{\dagger} +0.25\wp_{fiber} Z \rho Z^{\dagger},
\end{split}
\label{equ:fiber}
\end{equation}
and
\begin{equation}
\begin{split}
\varepsilon(\wp_{memory})& = (1-\wp_{memory}) I \rho I^{\dagger} + \wp_{memory} X \rho X^{\dagger},
\end{split}
\label{equ:memory}
\end{equation}

% 
% The probability $p_{fiber}$ and $p_{memory}$ are the function of length of fiber through which a qubit is transmitted\cite{Li2023} and the function of storage time of quantum memory where qubit is stored\cite{}, expressed as:

The probability $\wp_{fiber}$ and $\wp_{memory}$ are mathematically expressed as $\wp_{fiber} = 1- 10^{-\eta l/10}$  and $\wp_{memory} = 1- exp(-t_{w}/T)$, respectively.
%\begin{equation}
%p_{fiber} = 1- 10^{-\eta l/10} 
%\label{equ:p_fiber}
%\end{equation}
%and,
%\begin{equation}
%p_{memory} = 1- exp(-t_{w}/T)
%\label{equ:p_memory}
%\end{equation}

Here, $\eta$ and $l$ in $p_{fiber}$ represent the attenuation factor of optical fiber (in dB/km) and the corresponding length of qubit transmission, respectively. Given a single path from Fig. \ref{fig:archi} with total length $L$ and $\mathcal{M}$ edges, the edge length $l$ can be calculated as $l = L/\mathcal{M}$. Similarly, $t_{w}$ and $T$ in $\wp_{memory}$ are defined as the waiting time for a qubit in memory and the corresponding loss-related time constant that is defined by the material properties of the memory \cite{gera2020mhz}. Specifically,  the length $l$ and waiting time $t_{w}$ are  variables that play a critical role in calculating the edge cost. %These variables are essential for the design of the caching network protocol, as elaborated in the subsequent sections.

\subsection{Cost Metric Definition }
By employing the error model, this section aims to establish the definition of cost metrics for each edge, which includes both the time overhead in memory and the overhead of traversing the length of the fiber channel.
% Specifically, the incurred time overhead ($t_{w}$) is a linear combination of both entanglement swapping-related and queue delay factors, calculated as $t_{w} = t_{w-e} + t_{w-q}$. 
The theoretical analysis of the cost metric is performed on a given path in the graph denoted as $G=\{P, N, E, C\}$ with notations as defined in Sec.~\ref{subsec:network_architecture}.

% The goal of this section aims to develop a system model regarding to the time synchronization while establishing end-to-end entanglement with multiple quantum links of each path, and the corresponding performance calculation. As discussed in \textcolor{red}{Section XXX}, we consider each quantum path represented by a graph $G=\{N,E,C\}$, with $N$ as the nodes, $E$ as the edge and $C$ as the storage.  

\begin{figure}[b!]
\centering
\vspace*{-10pt}
\includegraphics[width=1.0\linewidth]{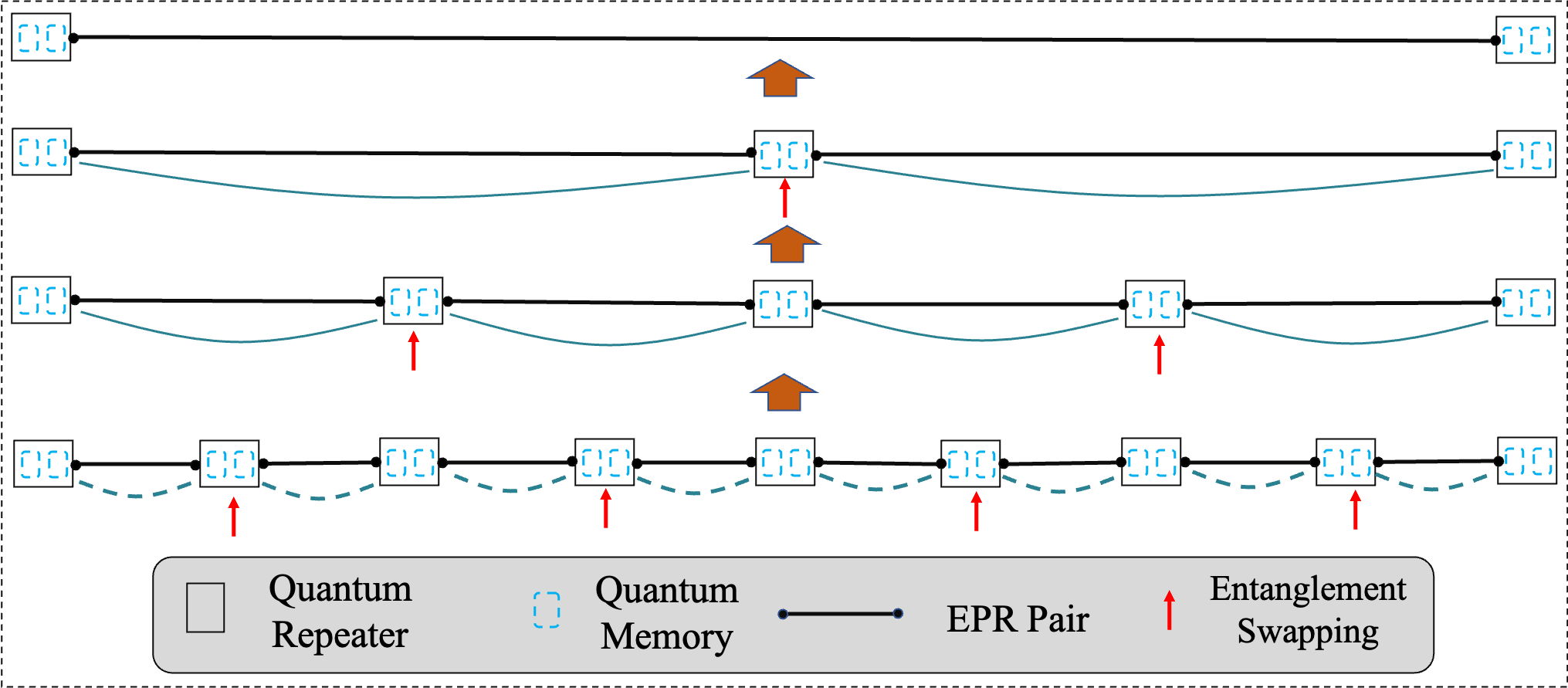}
\caption{Iterative entanglement swapping operations to establish long distance EPR pair. In this example, $\mathcal{J} = 3$ iterations are needed to establish end-to-end entangled pair of qubits. }
\vspace*{-10pt}
\label{fig:time_w_s}
\end{figure}

\noindent\textbf{Overhead of entanglement swapping:} In this section, we first analyze the overhead associated with the waiting time resulting from entanglement swapping in one path, denoted as $t_{w-e}$. In the following section, we obtain the time-related overhead due to waiting time in the queuing using queuing theory.  As depicted in Fig. \ref{fig:time_w_s}, the establishment of entangled Einstein-Podolsky-Rosen (EPR) pairs with longer distances involves sequential entanglement swapping operations with smaller distances, progressing from the bottom to the top.  
%In our analysis, consider a graph denoted as $G=\{P, N, E, C\}$ using the terms defined in Sec.~\ref{subsec:network_architecture}. 
We  specifically consider each path composed of $\mathcal{M}$ edges, and each edge is associated with one EPR pair. Notably, the shortest edge depicted in the diagram represents one of these $\mathcal{M}$ edges. The process of entanglement swapping, denoted by a red arrow, is performed by two repeaters in close proximity, with at least one intervening edge between them. These entanglement swapping events take place simultaneously. Following this, the entanglement swapping process operates iteratively on the new entangled pairs, ultimately leading to the establishment of the final end-to-end entangled pair. The qubits that are not involved in the entanglement swapping process are temporarily stored in the quantum memory. %This waiting time is calculated based on the length of the edge they correspond to. 
Using the approach presented in \cite{waiting_1}, for a path with $M$ edges, the total waiting time $t_{w-e}$ can be calculated as follows:
\begin{equation}
t_{w-e} = \sum_{j=1}^{\mathcal{J}}(2^{j-1}\frac{l}{c}).
\label{equ:time_w_e}
\end{equation}

Here, $c$ denotes the speed of light in the fiber, $l$ represents the distance of the shortest edge, and $\mathcal{J}$ signifies the number of iterations of the entanglement swapping operation in one path, for e.g., $\mathcal{J} = 3$ in Fig. \ref{fig:time_w_s}.
% where the details of the calculation $\mathcal{J}$ can be refereed in\cite{shchukin2019waiting}. 
During this process, the two qubits at both ends of a given path are  in a waiting state until all iterations are completed. The waiting time, as calculated in Eq. \eqref{equ:time_w_e}, is then employed as the time-related overhead for the entanglement swapping procedure. Secondly, the length-related overhead is solely determined by the fiber length given by  the edge length $l$. 
 
% \begin{figure}[b!]
% \centering
% \includegraphics[width=1.0\linewidth]{Figures/e-w.png}
% \caption{Iterative entanglement swapping operations to establish long distance EPR pair}
% \label{fig:time_w_s}
% \end{figure}

\noindent\textbf{Edge cost calculation:}  The edge cost is calculated by considering both %the time overhead and the length overhead factors, which take into account 
the effects of waiting time in memory and the fiber transmission length. This approach provides a comprehensive measure of the cost associated with the $m^{th}$ edge $e_m$. Specifically, the edge cost of the $m^{th}$ edge $e_m$ is denoted by $\mathcal{C}_{e_m}$ and is defined as the fidelity (represented as $f_m$ \cite{request_schedule}) loss of the associated qubit:
\begin{equation}
% \begin{split}
\mathcal{C}_{e_m} = 1-f_m = 1- Tr[\sqrt(\sqrt(\rho_m)\epsilon(\wp)_{m}\sqrt(\rho_m))]^2 ,
% \end{split}
\label{equ:edge_cost1}
\end{equation}
here, the $\rho_m$ and $\epsilon(\wp)_{m}$ are density matrix and the error operator of the $m^{th}$ edge, respectively, which are derived via Eq. \eqref{equ:erroioerator}  to \eqref{equ:memory}. 
Next, we consider one path with $\mathcal{M}$ edges. The total fidelity of the that path is calculated using the recursive function $\Theta(\mathcal{C}_{e_1}, \mathcal{C}_{e_2},..., \mathcal{C}_{e_\mathcal{M}})$. The total cost of the path is donated by $\mathcal{C}_T$ and equals the fidelity (represented as $f_{T}$) loss in one path. Formally:
\begin{equation}
\mathcal{C}_T = 1-f_{T}
= 1- \Theta(\mathcal{C}_{e_1}, \mathcal{C}_{e_2},..., \mathcal{C}_{e_\mathcal{M}} ) .
% \\&= 1- (\Theta(\mathcal{C}_{e_1}, \mathcal{C}_{e_2}) + \Theta(\mathcal{C}_{e_2}, \mathcal{C}_{e_3}) + \cdots + \Theta(\mathcal{C}_{e_{M-1}}, \mathcal{C}_{e_{M}}) )
\label{equ:total_cost}
\end{equation}

% $\Theta(f_{1}, f_{2},..., f_{M} )$ is a defined function that is used to calculate the total fidelity of one path with $m$ edges. 

The recursive function $\Theta(\mathcal{C}_{e_{1}}, ...,\mathcal{C}_{e_\mathcal{M}} )$ is formulated as:

\begin{equation}
\begin{split}
\Theta(\mathcal{C}_{e_{1}}, ...,\mathcal{C}_{e_\mathcal{M}} )  &= (1-\Theta(\mathcal{C}_{e_{1}}, ...,\mathcal{C}_{e_{\mathcal{M}-1}} ))(1-\mathcal{C}_{e_{\mathcal{M}}})\\
&+\frac{\Theta(\mathcal{C}_{e_{1}}, ...,\mathcal{C}_{e_{\mathcal{M}-1}} )\mathcal{C}_{e_{\mathcal{M}}}}{3},
\end{split}
\label{equ:F_final}
\end{equation}
where $\Theta(\mathcal{C}_{e_{1}}) = \mathcal{C}_{e_{1}}$, calculated from Eq.~\eqref{equ:edge_cost1}.

\subsection{System Model of \name}
This section explains the system model of \name ~involving qubit queuing within the quantum memory. 
% Subsequently, the section formulates a system performance evaluation benchmark that encompasses objectives and constraints pertinent to the design of the caching network protocol. By outlining these aspects, we lay the foundation for a comprehensive analysis of the caching network's effectiveness in managing qubit queuing and optimizing the overall quantum communication system performance.

\noindent\textbf{Queuing model in a caching network:} 
% Qubit repetition method facilitates the transfer of  one logicalal qubits encoded by $K$ physical qubits.
% therefore enhancing the accuracy of transmitted qubit data information.
% Quantum memory enables parallel entanglement swapping operations for each transmitted logicalal qubit, resulting in an increased logicalal qubit transmission rate. However, this enhancement comes at the expense of additional losses incurred due to the waiting time experienced in the memory while queuing for detection by the receiver. A caching network protocol via queuing model is designed to exemplify the trade-off between the transmission rate and the supplementary cost arising from additional queuing delay. 
% Additionally, the designed protocol  
% In this model, variation of the the number of memories in each nodes is considered as the variables,  but the variation of number of nodes will not be considered. Additionally, the number of quantum memories in each nodes in each path are considered as same.
As depicted in Fig.~\ref{fig:queue_1}, a single logical qubit is encoded by $K$ physical qubits, which are broadcasted through $K$ viable paths interconnected by quantum repeaters, ultimately reaching the receiver.  Each repeater is equipped with $\mathcal{I}$ quantum memory units, enabling the simultaneous processing of the necessary entanglement swapping operations for the transmission of $\mathcal{I}$ logical qubits.  
During the processing of transmitting $\mathcal{I}$ logical qubits, the quantum information of each logical qubit is received at the receiver from $K$ path. The quantum information of non-received logical qubits is stored in the queue through the utilization of quantum memory. 
We refer to this process of receiving quantum information of logical qubit at receiver as \textit{serving}, with a serving rate denoted as $\gamma$. While serving is underway, the remaining logical qubits stored in the quantum memory units are placed in a queue, which we refer to  as \textit{queuing}. Such queuing introduces an additional waiting time known as queuing delay, denoted as $t_{w-q}$. 
 % These queued qubit will be called and being served once the previous serving process is finished.
Furthermore, while one logical qubit is serviced, the new logical qubits are continuously generated and distributed to the memory, irrespective of whether the memory is empty or already occupied. Analogous to the classical data network, if the quantum memory is occupied, new logical qubit is dropped. We consider the process of the newly generated logical qubits arriving in the memory as the \textit{arriving}, with \textit{arrival rate}  denoted as $\lambda$. While we model  the logical qubit queuing process in the caching network in three steps: \textit{serving}, \textit{queuing}, and \textit{arriving}, the following challenges need to be addressed:

\begin{figure}[t!]
\centering
\includegraphics[width=1.0\linewidth]{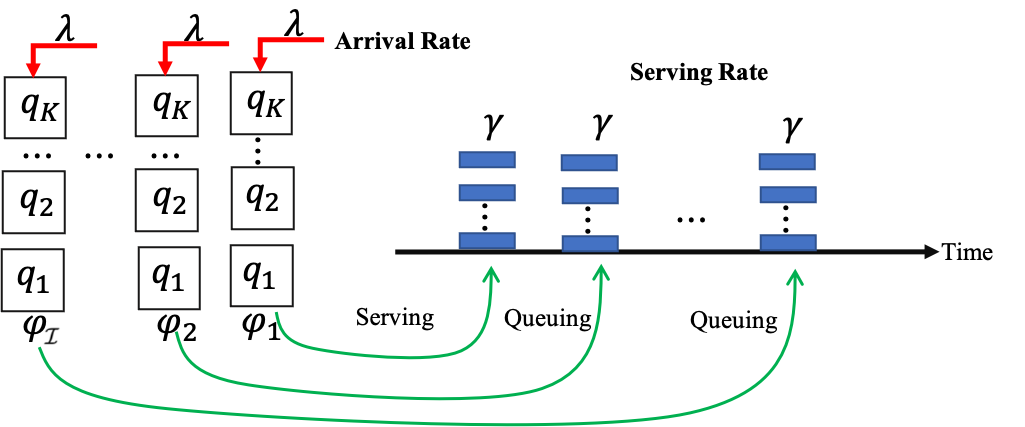}
\caption{Queuing model of qubits in caching network, logical qubits are served in a sequential queue with arrival rate $\lambda$, serving rate $\gamma$.}
\vspace*{-15pt}
\label{fig:queue_1}
\end{figure}

\noindent\textbf{Challenge 1}: The calculation of the waiting time, denoted as $t_{w-q}$, in quantum memory involves the queuing of logical qubit transmissions, taking into account their distribution with respect to arrival rate, serving rate, and the number of available quantum memory. 

\noindent\textbf{Challenge 2}: During the queuing process, qubits are stored in the memory for an extended duration, resulting in a gradual degradation of qubit fidelity and an exponential increase in the edge cost. Although a higher number of memory units may offer higher rates, it also entails longer waiting times per qubit (higher edge cost). As a result, a careful balance must be struck between the transmission rate and the associated cost to optimize the system's performance.

% \noindent \textbf{Solution 1}: A new metric that evaluates the performance of the system is demanded, and the corresponding optimization algorithm is required.

\noindent \textbf{Challenge 3}: The disparity between \textit{serving} and \textit{arrival}  rates leads to imbalances and inefficiencies within the queuing process, affecting the overall transmission and waiting times, which pose challenges in accurately evaluating the system's performance and managing available resources optimally.

% \noindent \textbf{Solution 2}: We integrate the success probability of into the arrival rate and serving rate at each round, and formulate the arrival rate and serving rate as the Poisson distribution\cite{vardoyan2019stochastic} with the average value of \textcolor{red}{XXX} and \textcolor{red}{XXX}

% \noindent \textbf{Challenge 3}: The quantum error correction method require encode $k$ physical qubits into one logical qubit, and 
% $k$ transmission path are required to  distribute these qubit. However, the different length among the $k$ path challenges the synchronization of serving of all qubits and the optimization of system performance.  

% \noindent \textbf{Challenge 4}:  The stochastic nature associated with the success probability of entanglement swapping operations poses challenges in predicting outcomes with certainty. This uncertainty introduces complexities when assessing the overall efficiency and reliability of the quantum communication system. 

% \noindent \textbf{Solution 3}: For each path, we leverage one queue protocol, and then evaluate the system performance with the combination of multiple path queuing protocols  and the leverage the \textcolor{red}{XXXX} based  optimization algorithm. 
In $\name$, we overcome Challenge 1 by using results from queuing theory and Challenges 2-3 by formulating an optimization problem in Section III.F.

% \subsection{Problem formulation and $\name$ by Queuing Theory}
\subsection{Wait Time in Queuing Calculation in $\name$}

This section formulates the problem and proposes solutions to the challenges encountered in the caching network design by leveraging queuing theory.
% Specifically, we utilize an embedded Markov chain for the analysis of waiting time in queues.

\begin{figure}[t!]
\centering
\subfloat[Queuing model]{
\centering
\includegraphics[width=0.85\linewidth]{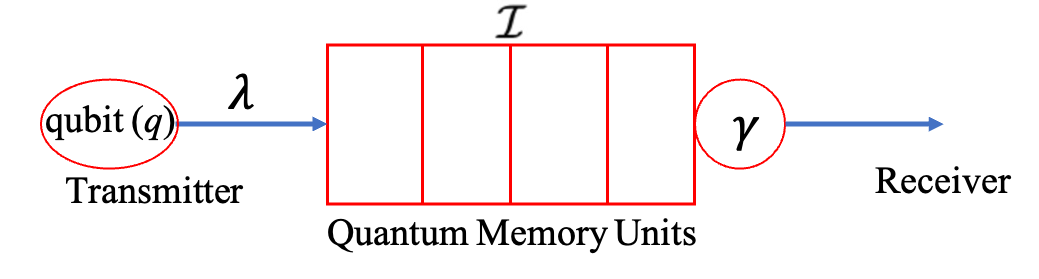}
\label{fig:queue_model}}
\vspace*{-10pt}

\subfloat[Transition diagram]{
\centering
\includegraphics[width=0.85\linewidth]{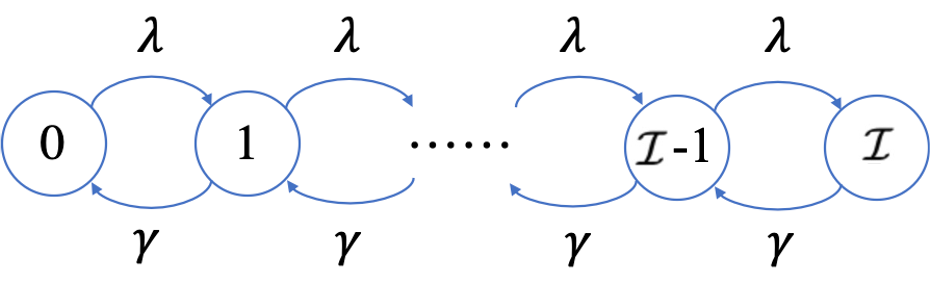}
 \label{fig:markov_chain_trans}}

\caption{The $M/D/1/\mathcal{I}$ queue, modeling queuing process over one path. Each repeater on the path is equipped with $\mathcal{I}$ quantum memory units. Qubits are generated by the transmitter following Poisson process with rate $\lambda$, and are served by the receiver deterministically with rate~$\gamma$.}
\vspace*{-10pt}
% \label{fig:markov_chain}
\end{figure}

\noindent\textbf{Waiting time analysis in queuing:} 
The above-mentioned challenges pose constraints to getting an explicit solution on the waiting time analysis during the queuing process. To deal with these challenges, we follow the finite-capacity $M/D/1/\mathcal{I}$ queuing model. In short, $M/D/1/\mathcal{I}$ queuing model is compatible with the finite capacity queuing system and the FCFS (First Come First Served) discipline. With $M/D/1/\mathcal{I}$ queuing model, given any path we consider the arrival rate of a physical qubit (same for encoded logical qubit) from the transmitter as {\em Poisson distribution} \cite{vardoyan2019stochastic}, and the serving rate at the receiver for each physical qubit as deterministic, which are depicted as $\lambda$ and $\gamma$ in the simplified queuing model in Fig. \ref{fig:queue_model}. Mathematically, the waiting time analysis of each queued physical qubit in the memory includes two stages. At the first stage, the embedded Markov chain associated $M/D/1/\mathcal{I}$ queuing model\textcolor{black}{\cite{lovas2021markov}} is considered, which derives the probability distribution of the number of physical qubits queued in the system based on the Markov chain transform shown in Fig.~\ref{fig:markov_chain_trans}.  At the second stage, by leveraging the probability distribution and {\em Little's Law}, the mean number of physical qubits in quantum memory and associated mean waiting time of the $M/D/1/\mathcal{I}$ queuing is obtained. Therefore, following \cite{brun2000analytical}, given the proposed quantum caching network with arrival rate $\lambda$, serving rate $\gamma$ and the number of memory units $\mathcal{I}$, the average waiting time $t_{w-q}$ experienced in a repeater while being queued can be summarized as: 
% {\color{red} why sudden citation here?}
\begin{equation}
    t_{w-q} = (\mathcal{I}-1-\frac{\sum_{i=0}^{\mathcal{I}-1}b_{i}-\mathcal{I}}{\varrho b_{\mathcal{I}-1}})\frac{1}{\gamma},
\label{equ:waiting_t_wq}
\end{equation}
\begin{equation}
    b_{\mathcal{I}} = \sum_{i=1}^{\mathcal{I}}\frac{(-1)^{i}}{i!}(\mathcal{I-i})^{i}e^{(\mathcal{I}-i)\varrho}\varrho^{i},
\label{equ:b_cofe}
\end{equation}
where $\mathcal{I}$ is the total number of quantum memory units of each repeater, $i$ represents the $i^{th}$ quantum memory unit of all memory units, $\varrho = \lambda / \gamma$ is the utilization factor of quantum memory, $b_{\mathcal{I}}$ is the coefficient of the probability transition matrix of the embedded Markov chain with $z-transform$ \cite{brun2000analytical}.  

\noindent\textbf{Time overhead variation:} 
This section describes the cost metric in terms of time overhead and the corresponding edge cost. Due to the extra waiting time in the queuing, the total time overhead for each qubit transmission in the system is the superposition of the time overhead in entanglement swapping and the time overhead during the queuing, expressed as: 
\begin{equation}
% \begin{split}
     t_{w} = t_{w-e} + t_{w-q},
     % t_{w-e}& = \sum_{j=1}^{\mathcal{J}}(2^{j}\frac{l}{c})\\
     % t_{w-q}& = (\mathcal{I}-1-\frac{\sum_{i=0}^{\mathcal{I}-1}b_{i}-\mathcal{I}}{\varrho b_{\mathcal{I}-1}})\frac{1}{\gamma}
% \end{split}
\label{equ:t_time_overhead}
\end{equation}
where $t_{w-e}$ and  $t_{w-e}$ are  defined in the Eqs.~\eqref{equ:time_w_e} and \eqref{equ:waiting_t_wq}, respectively. 
% The corresponding parameters involved in this equation can be found in Eq.~\ref{equ:time_w_e}. 
It is observed that the calculation of edge cost $\mathcal C_{e_{m}}$ and total cost of each path $\mathcal{C_{T}}$ can be accordingly modified with the total time overhead represented in Eqs.~\eqref{equ:edge_cost1} and \eqref{equ:total_cost}.

% \subsection{Evaluation metric of \name}
\subsection{Problem Formulation of $\name$}
This section formulates the optimization problem as logical qubit transmission rate associated with decoding accuracy in the proposed caching network.

\begin{figure}[h!]
\centering
\vspace*{-10pt}
\includegraphics[width=0.9\linewidth]{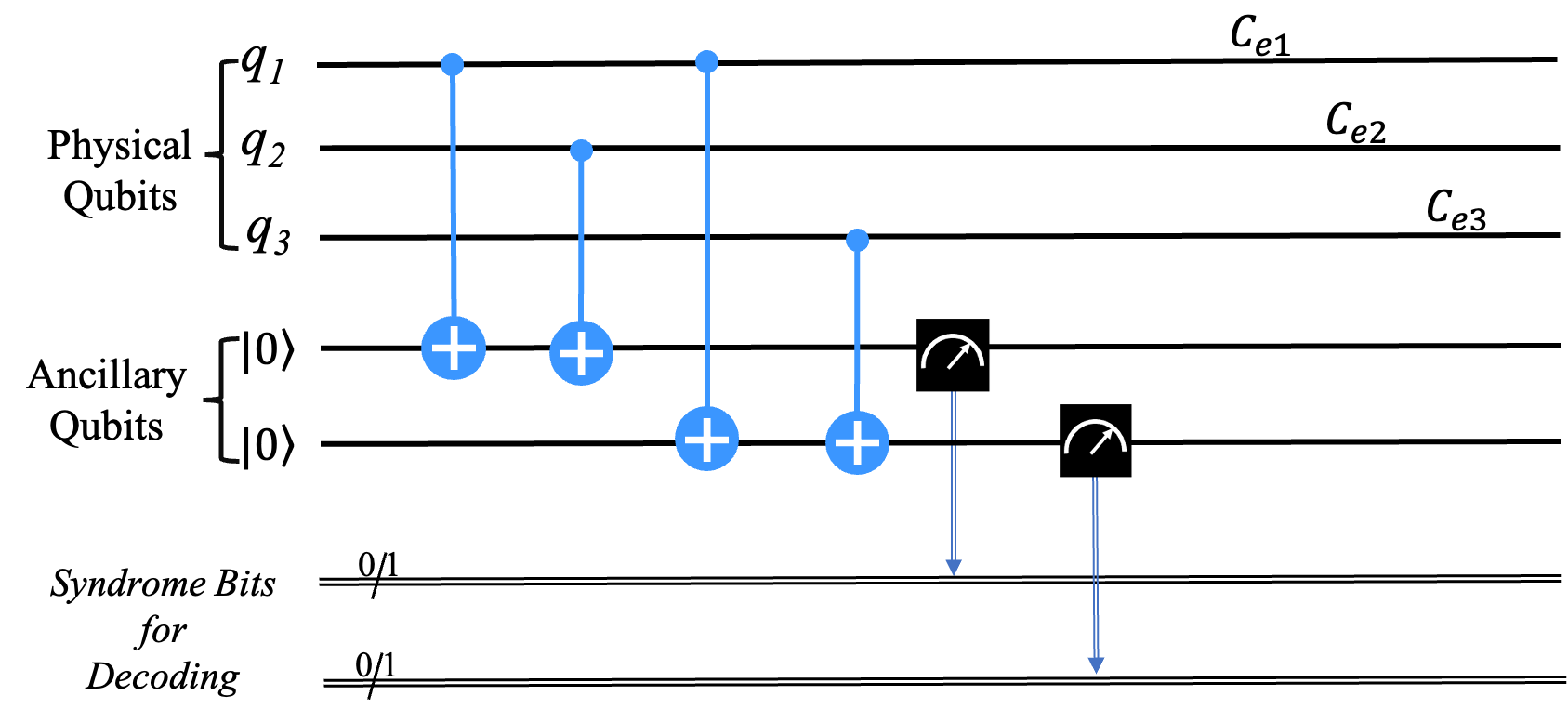}
\caption{Quantum circuit schematic with encoded physical qubits, such as $ K=3$ in this case. Syndrome bits are obtained with ancillary qubits without destroying the transmitted physical qubits.}
\vspace*{-10pt}
\label{fig:q_circuit}
\end{figure}

\noindent\textbf{Accuracy of logical qubit decoding:} We assume that the receiver has a  transducer that is capable of converting the photonic qubit as quantum computing based qubit for decoding the received logical qubit. We design the quantum circuit for decoding the logical qubit via analyzing the syndrome bits measured by ancillary qubits, and one such example is given in Fig.~\ref{fig:q_circuit}. Ancillary qubits are used to extract information about errors in transmitted qubits through entanglement, and the syndrome bits are obtained by measuring the ancillary qubits to identify the presence and location of errors without directly measuring the transmitted qubits themselves. Here, the syndrome bits capture the information about qubit errors that are caused by the associated costs, such as $\mathcal{C}_{e_1}, \mathcal{C}_{e_2}, \text{ and } \mathcal{C}_{e_3}$ in Fig.~\ref{fig:q_circuit}. Two different decoding methods are utilized to obtain the accuracy of the received logical qubit over the proposed caching network. In particular, these two decoding methods are direct maximum weight matching and lookup table method \cite{das2022lilliput}. The corresponding error probability of each logical qubit is denoted as $\mathcal{P^{W}}$ and $\mathcal{P^{L}}$. The designed quantum circuit for decoding the logical qubit is depicted in Fig. \ref{fig:q_circuit}.

% The ancillary qubits will not destroy the physical qubits.

% From Fig. \ref{fig:q_circuit}, it is observed that the obtained syndrome bits from related to number of physical qubits ($K$) used for encoding one logicalal bit, and the cost ($\mathcal{C}_{e_{m}}$) of each physical qubit during the transmission associated with total time overhead and the length overhead. 

Therefore, $\mathcal{P^{W}}$ and $\mathcal{P^{L}}$ will be modified  as $\mathcal{P^{W}}(K, \mathcal{C}_T)$ and $\mathcal{P^{L}}(K, \mathcal{C}_T)$. The simulation results provide the proof of the variation of $\mathcal{P^{W}}(K, \mathcal{C}_T)$ and $\mathcal{P^{L}}(K, \mathcal{C}_T)$ with the change of number of physical qubits and the corresponding costs. 

\noindent\textbf{Ideal logical qubit transmission rate:} In the proposed caching network, we define the 
logical qubit transmission rate as the reverse of time overhead spend in the system, expressed as: 
\begin{equation}
R_{logic} =1 / t_{w} = 1/ (t_{w-e}+ t_{w-q}).
\label{equ:rate_logic}
\end{equation}

Given the imperfect decoding of the logical qubit, the ideal logical qubit transmission rate is calculated as: 
\begin{equation}
\begin{split}
R_{logic}^{\mathcal{W}}&=\frac{1}{t_{w}}(1-\mathcal{P^{W}}(K, \mathcal{C}_T) ) = \frac{1}{t_{w-e}+ t_{w-q}} (1-\mathcal{P^{W}}(K, \mathcal{C}_T) ),\\
R_{logic}^{L}&=\frac{1}{t_{w}}(1-\mathcal{P^{L}}(K, \mathcal{C}_T) ) = \frac{1}{t_{w-e}+ t_{w-q}} (1-\mathcal{P^{L}}(K, \mathcal{C}_T)).
\end{split}
\label{equ:corrected_rate}
\end{equation}

In particular, \name~solves the following optimization problem to obtain the ideal qubit transmission rate: 
\begin{subequations}
\label{eq:opt_twin_selection}
\begin{align}
% \begin{gather}
    \underset{l,\mathcal{J},K,\mathcal{I}}{max} &~~\frac{1}{t_{w-e}+ t_{w-q}}(1-\mathcal{P}^{S}(K, \mathcal{C}_T)),\\
    \text{Given} ~~&     t_{w-e} = \sum_{j=1}^{\mathcal{J}}(2^{j}\frac{l}{c}),\\
     &t_{w-q} = (\mathcal{I}-1-\frac{\sum_{i=0}^{\mathcal{I}-1}b_{i}-\mathcal{I}}{\varrho b_{\mathcal{I}-1}})\frac{1}{\gamma},\\                 
    \text{s.t.}~~& \mathcal{P}^{S}(K, \mathcal{C}_T) \in \{ \mathcal{P^{M}}(K, \mathcal{C}_T), \mathcal{P^{L}}(K, \mathcal{C}_T)\},\\
    & \mathcal{C}_T > Thres_1,
\end{align}
\end{subequations}
where $Thres_1$ is a threshold of cost in each path defined by the users following their requirements.

%% file: section/result.tex
\section{Results and Discussions}
\label{sec:exp}
\vspace*{-3pt}
In this section, we evaluate $\name$ using the theoretical model described in Sec.~\ref{sec:systemdesign}. First, we obtain the waiting time $t_{w-q}$ caused by queuing delay. Second, we plot the fidelity of the physical qubit in a given path. Third, we calculate the probability of error for a logical qubit decoding. Finally, $\name$ sets  the appropriate logical qubit transmission rate, and the improvement in the corresponding error correction accuracy is shown for the overall system evaluation.  All of the parameters used for the evaluation of the proposed \name~caching network are listed in Table~\ref{tab:parameter}.

\begin{table}[h!]
\vspace{2mm}
\centering
\scalebox{0.9}{
\begin{tabular}{|| p{0.1\linewidth} | p{0.35\linewidth} | p{0.40\linewidth}|| }
\hline
 \textbf{Notation} & \textbf{Meaning} & \textbf{Value} \\
 \hline \hline
  $c$ & Light speed in fiber &  $2\times 10^{8}$ m/s \\ 
\hline
  $L$ & Edge length & 80 km, 120 km \\ 
\hline
  $\eta$ & Attenuation Factor & 0.2 dB/km  \\ 
\hline
  $K$ & Number of paths (or physical qubits) & Variables\\ 
\hline
  $\mathcal{M}-1$ & \# Quantum repeater & Variables\\ 
\hline
  $\mathcal{I}$ & \#  Quantum memory units & Variables\\ 
\hline
 $\lambda$ & Arrival rate & [$0.05$, $0.2$,$ 0.5$,
 $1.2$ ] MHz \\ 
\hline
 $ \gamma $& Serving rate & [$0.02$, $0.025$, $0.05$, $0.1$, $4.41$]  MHz \\ 
\hline
  $t_w$ & Total time overhead & Eq. \eqref{equ:t_time_overhead}\\ 
\hline
  $t_{w-e}$ & Time overhead caused by entanglement swapping  & Eq. \eqref{equ:time_w_e}\\ 
\hline
  $t_{w-q}$ & Time overhead caused by queuing delay & Eq. \eqref{equ:waiting_t_wq}\\ 
\hline
\end{tabular}}
\caption{Parameters used for the evaluation of proposed \name.} 
\vspace*{-10pt}
\label{tab:parameter}
\end{table}

\begin{figure}[h!]
\centering
 \subfloat[]{
\includegraphics[width=0.48\linewidth]{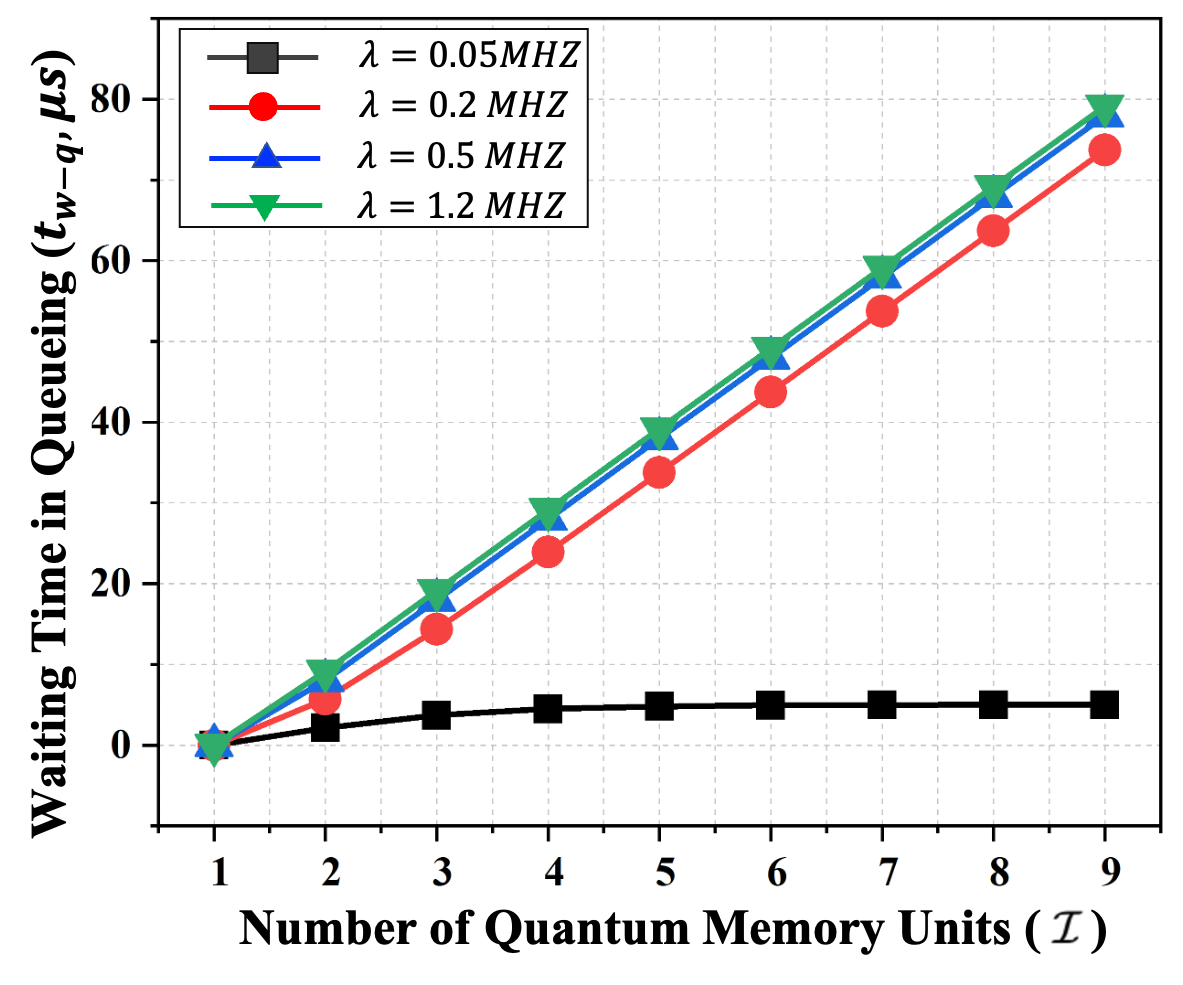}
% \caption{\textcolor{red}{Waiting time in Queue given entering rate $\lambda \in[0.05, 0.2, 0.5, 1.2 \times 10^{6} Hz]$ and serving rate $\gamma = 0.5\times 10^{6} Hz$ }. }
\label{subfig:queuing_time_1}}
 \subfloat[]{
 \includegraphics[width=0.48\linewidth]{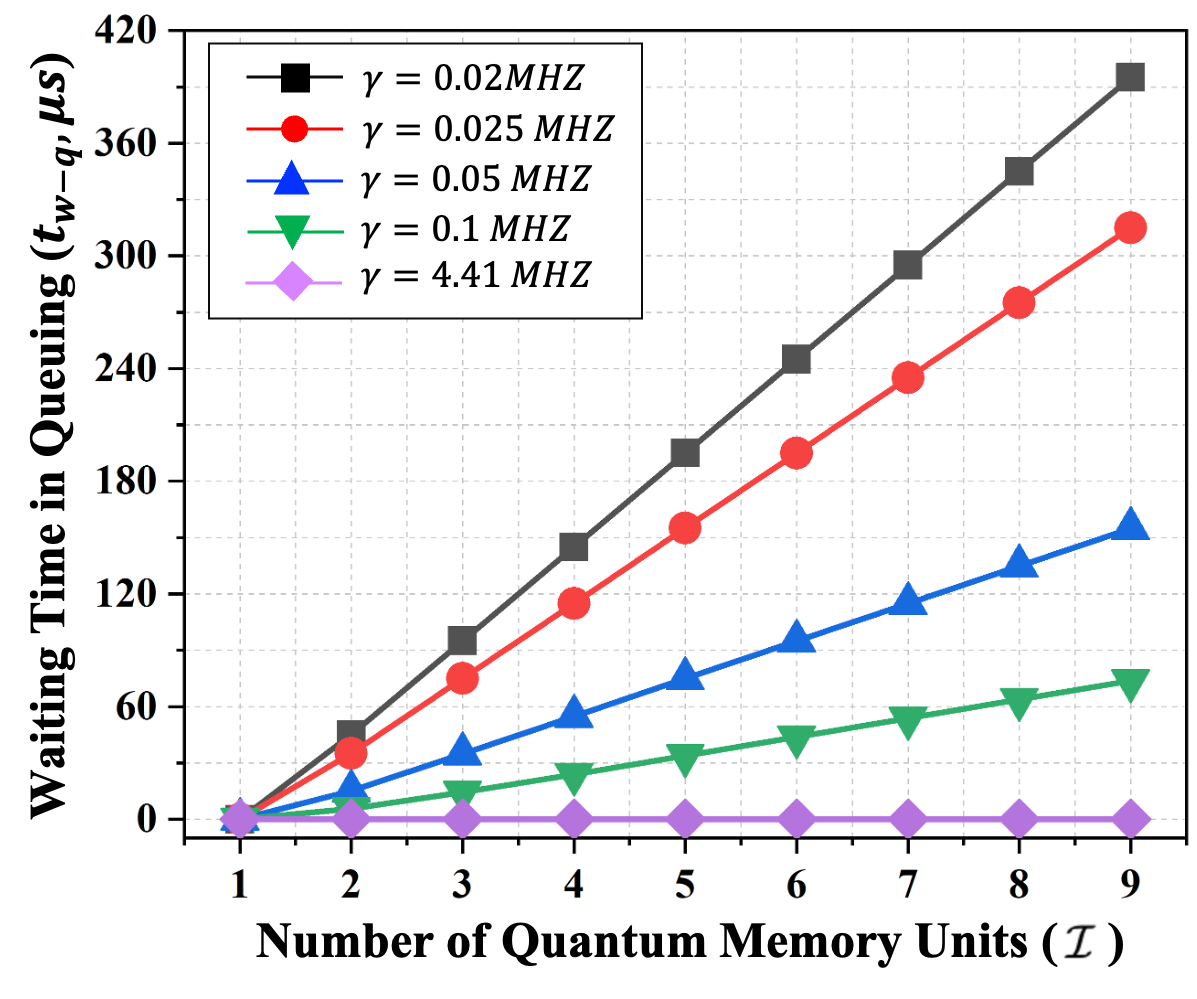}
% \caption{\textcolor{red}{Waiting time in Queue given entering rate $\lambda = 0.2 \times 10^{6} Hz]$ and serving rate $\gamma \in [0.02, 0.025, 0.05, 0.1, 4.41 \times 10^{6} Hz$}. }
 \label{subfig:queuing_time_2}}
 \caption{Waiting time $t_{w-q}$ analysis in queuing (a) arrival rate $\lambda \in[0.05, 0.2, 0.5, 1.2]$ MHz and serving rate $\gamma =  0.1$ MHz; (b) arrival rate $\lambda = 0.2$ MHz and serving rate $\gamma \in [0.02, 0.025, 0.05, 0.1, 4.41]$ MHz. Waiting time in queuing $t_{w-q}$ demonstrates a linear increase with a greater number of quantum memory units, also exhibiting varying rates of increase depending on the specific arrival and serving rates involved. }
 \vspace*{-13pt}
 \label{fig:queue_time}
\end{figure}

% \begin{figure}[h!]
% \begin{subfigure}{0.48\linewidth}
%   \centering
% \includegraphics[width=1\linewidth]{Figures/gamma_0.1_1.png}
% % \caption{\textcolor{red}{Waiting time in Queue given entering rate $\lambda \in[0.05, 0.2, 0.5, 1.2 \times 10^{6} Hz]$ and serving rate $\gamma = 0.5\times 10^{6} Hz$ }. }
% \label{subfig:queuing_time_1}
% \end{subfigure}
% \begin{subfigure}{0.48\linewidth}
%   \centering
%  \includegraphics[width=1\linewidth]{Figures/lambda_0.2.png}
% % \caption{\textcolor{red}{Waiting time in Queue given entering rate $\lambda = 0.2 \times 10^{6} Hz]$ and serving rate $\gamma \in [0.02, 0.025, 0.05, 0.1, 4.41 \times 10^{6} Hz$}. }
%  \label{subfig:queuing_time_2}
%  \end{subfigure}
%  \caption{Waiting time $t_{w-q}$ analysis in queuing (a) arrival rate $\lambda \in[0.05, 0.2, 0.5, 1.2]$ MHz and serving rate $\gamma =  0.1$ MHz; (b) arrival rate $\lambda = 0.2$ MHz and serving rate $\gamma \in [0.02, 0.025, 0.05, 0.1, 4.41]$ MHz. Waiting time in queuing $t_w-q$ demonstrates a linear increase with a greater number of quantum memory units, also exhibiting varying rates of increase depending on the specific arrival and serving rates involved. }
%  \vspace*{-13pt}
%  \label{fig:queue_time}
% \end{figure}

\subsection{Waiting time Analysis in Queuing} 
\vspace*{-3pt}
In this section, we study the waiting time introduced by queuing delay in a repeater. Since the different metrics of waiting time in queuing, the number of quantum memory units, arrival rate, and serving rate of the caching network have mutual dependencies, we examine their correlation patterns as derived by Eq. \eqref{equ:waiting_t_wq} and demonstrated in Fig. \ref{subfig:queuing_time_1} and \ref{subfig:queuing_time_2}, respectively. In Fig. \ref{subfig:queuing_time_1}, we consider the serving rate as 0.1 MHz. The observation indicates that the waiting time $t_{w-q}$ in queuing progressively increases with an increase in the number of memory units, irrespective of the arrival rate.Furthermore, $t_{w-1}$ demonstrates a significant disparity between the arrival rates $\lambda = [0.2, 0.5, 1.2]$ MHz, as compared to $\lambda = 0.05$ MHz. These findings indicate a scenario in which an arrival rate much lower than the serving rate leads to reduced effects on the increase of waiting time in queuing, albeit with the trade-off of a lower transmission rate.
% original:
% Additionally, $t_{w-q}$ shows relatively minor differences for arrival rates $\lambda = 0.5$ MHz and $\lambda = 0.2$ MHz, as well as for $\lambda = 1.2$ MHz and $\lambda = 0.05$ MHz. However, a substantial difference in waiting time is observed between $\lambda = 1.2$ MHz and $\lambda = 0.2$ MHz. 
% These results indicate that as the arrival rate approaches the serving rate, the waiting time significantly increases.
Moreover, the presence of a greater number of quantum memories in the quantum repeater prolongs the waiting time in queuing. In Fig. \ref{subfig:queuing_time_2}, we consider the arrival rate as $0.2$ MHz.  The waiting time in queuing rises rapidly for smaller $\gamma$ values, i.e., the arrival rates $\gamma = 0.02$ MHz and $\gamma = 0.05$ MHz.  Intuitively, we observe a linear increase in the waiting time in queuing as the number of memory units increase. 
\begin{observation}
\vspace*{-1pt}
The arrival rate, serving rate, and the number of memory units all contribute to the waiting time in queuing, which provides insights on how to  dynamically adjust waiting time overhead in queuing for any practical system (detailed in Figs. \ref{subfig:queuing_time_1} and \ref{subfig:queuing_time_2}, validates Contribution 1).
\vspace*{-3pt}
\end{observation}

\begin{figure}[ht!]
\centering
\vspace*{-10pt}
\subfloat[]{
\includegraphics[width=0.7\linewidth]{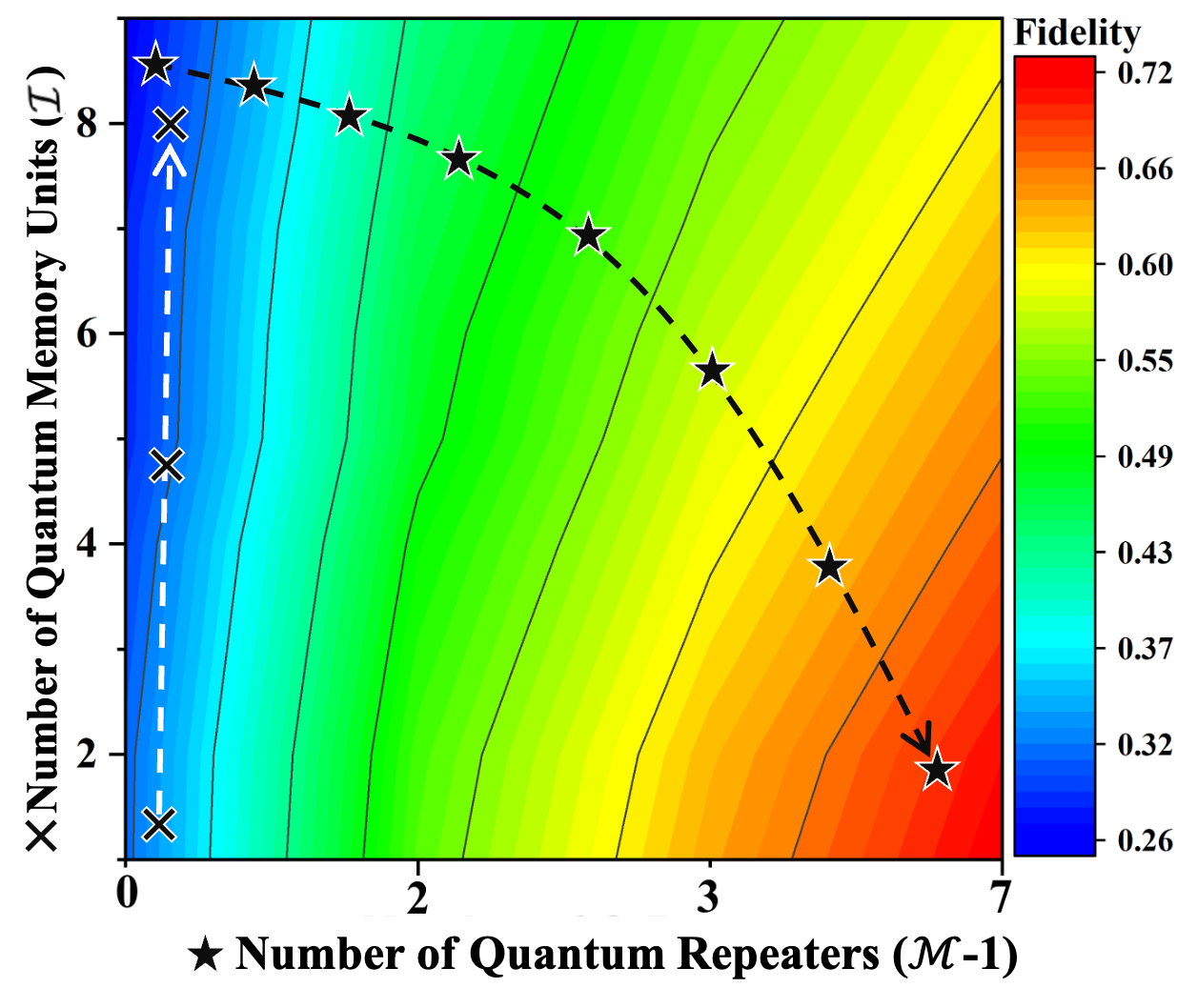}
\label{subfig:fidelity_80km}}

\vspace*{-0pt}
\subfloat[]{
\includegraphics[width=0.7\linewidth]{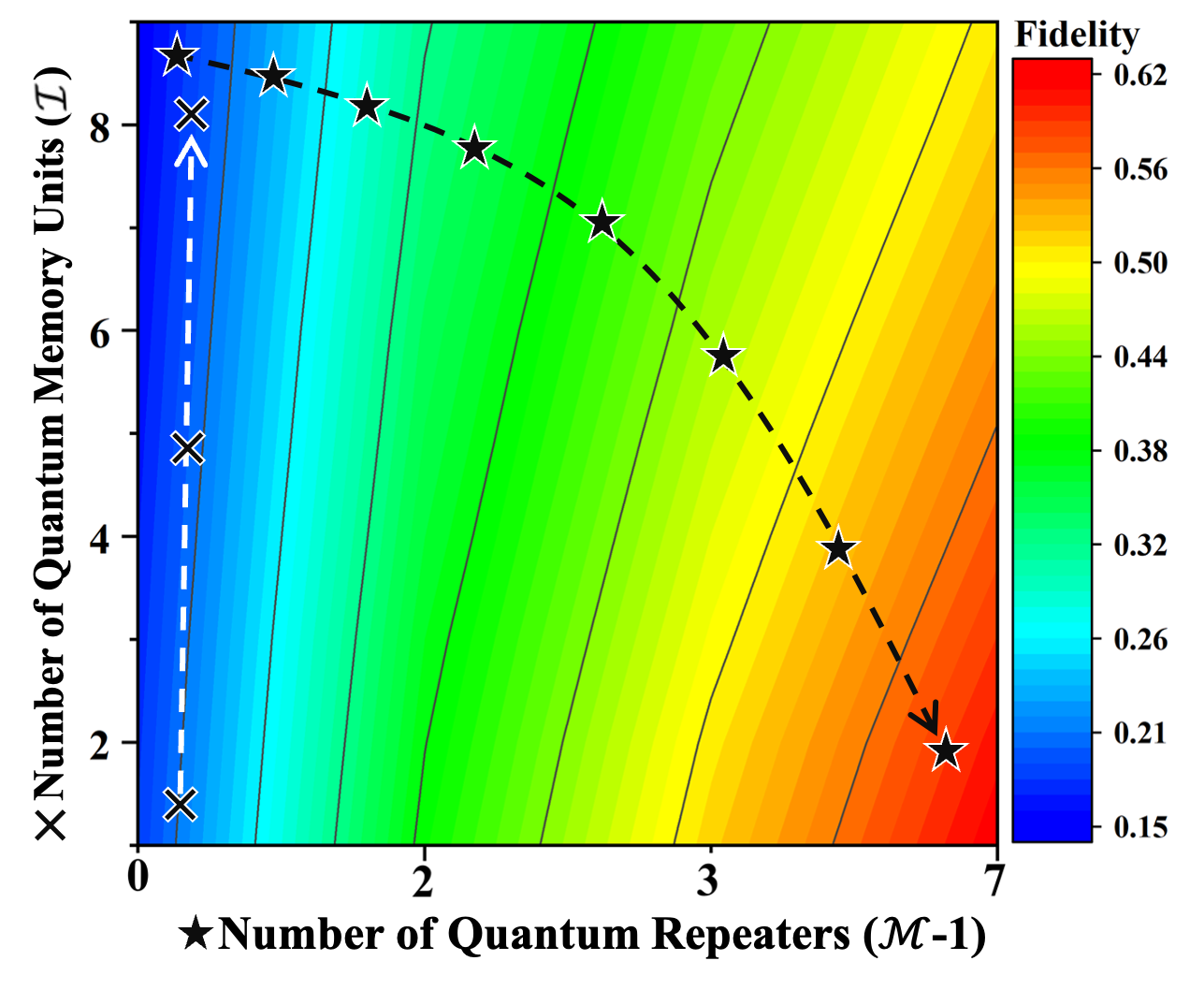}
\label{subfig:fidelity_120km}}
\vspace*{-1pt}
\caption{Fidelity change along with different number of quantum repeaters, memory units and path lengths: (a) $L = 80$ km and (b) $L = 120$ km. The fidelity demonstrate {\em increase trend} with the increasing $\mathcal{M}$, but {\em decrease trend} with increasing $\mathcal{I}$, indicated by and star and cross symbol, respectively.      }
\vspace*{-15pt}
\label{fig:fidelity}
\end{figure}

\subsection{Fidelity Analysis of Transmitted Qubits through one Path}
% \vspace*{-3pt}
We quantify the fidelity of a given qubit as a function of different numbers of memory units and repeaters. The fidelity calculation accounts for the temporal overhead attributed to entanglement swapping and queuing delay, alongside the loss due to fiber transmissions as given by the graph edge. 

% calculated using Eq.~\eqref{equ:total_cost}, taking into account the waiting time in the queue as demonstrated in Fig. \ref{fig:queue_time}. The analysis of EPR pair fidelity makes it evident how well the qubit states are preserved during transmission, providing insights into the data integrity.
% Moreover, the fidelity also serves as a manifestation of the level of total cost associated with each path.
Fig. \ref{subfig:fidelity_80km} and Fig. \ref{subfig:fidelity_120km} show how the fidelity of transmitted qubits vary  through a given path. The results presented herein are based on a random selection of arrival rates of $\lambda = 0.2$MHz with $\gamma = 0.025$MHz. We note that the fidelity trend remains consistent across various selections, demonstrating its generic nature. The findings from this set of simulations are used  Secs.~\ref{subsec:analysis_correct} and \ref{subsec:logical_qubit}. 
% The results obtained in this work use the same selection of the entering rate and serving rate.
With number of quantum memory units being constant, both figures exhibit a gradual increase in fidelity as the number of quantum repeaters involved in a single path increases, irrespective of whether the path length is $80$ km or $120$ km. This is a consequence of effectively reducing the loss due to traversing the fiber through the inclusion of additional quantum repeaters in the path. Despite the consequent increase in time overhead resulting from multiple entanglement swapping operations, the overall reduction in the loss due to traversal on shorter paths ultimately benefits the network. Furthermore, these results reveal that the inclusion of quantum memory leads to a decrease in the calculated fidelity, even when the number of quantum repeaters remains constant and irrespective of the path length being $80$ km or $120$ km. This decrease can be attributed to the additional time overhead caused by the prolonged waiting time in the quantum memory queuing. 
\begin{observation}
\vspace*{-5pt}
Different number of quantum repeaters and quantum memory units impact the final fidelity of the received qubit from different perspectives, 
% , time overhead in the queue due to quantum memory, and both time/length overhead in entanglement swapping operation in the quantum repeater.
which underscores the significance of tuning these parameters to achieve desired performance levels (details in Fig. \ref{subfig:fidelity_80km} and \ref{subfig:fidelity_120km}, validates Contribution 2).
\vspace*{-2pt}
\end{observation}

\begin{figure*}[ht!]
\centering
\subfloat[]{
\includegraphics[width=0.5\linewidth]{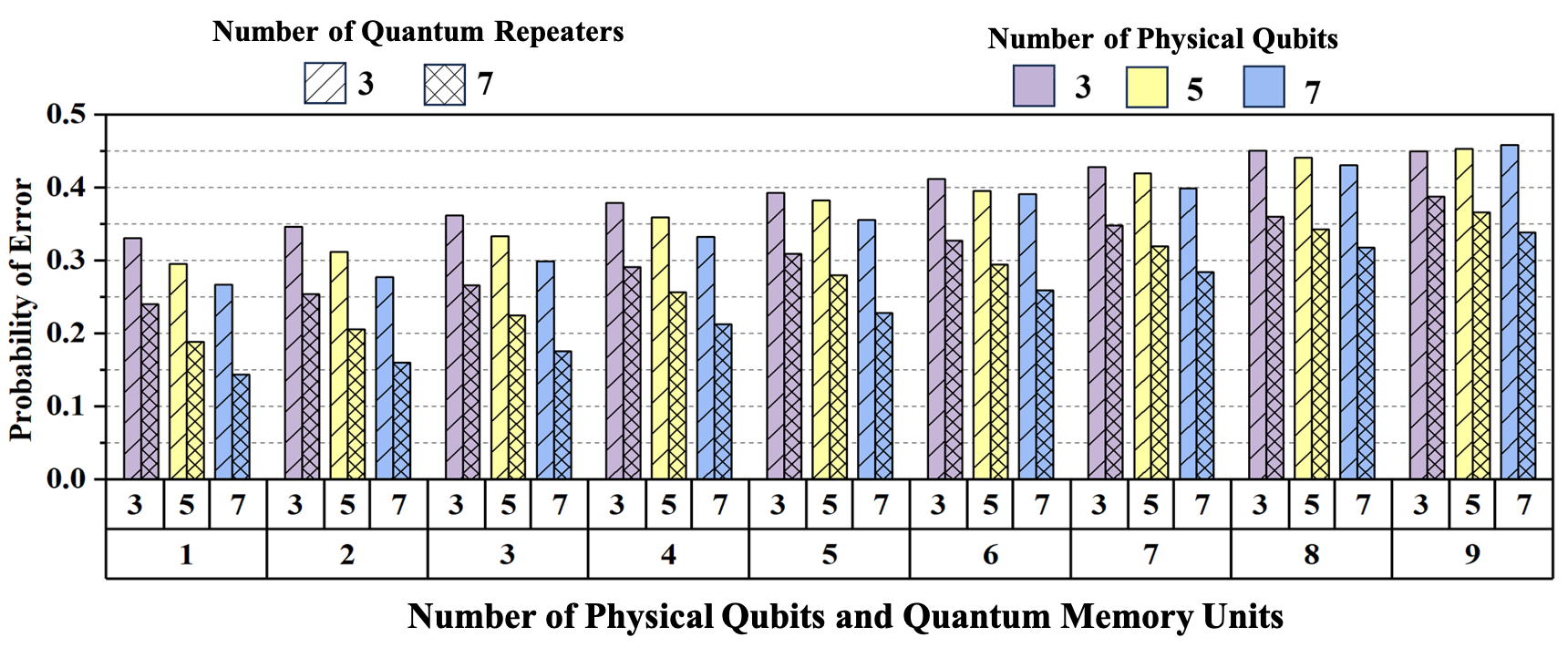}
\label{fig:error_1}}
\hspace*{-10pt}
\subfloat[]{
\includegraphics[width=0.5\linewidth]{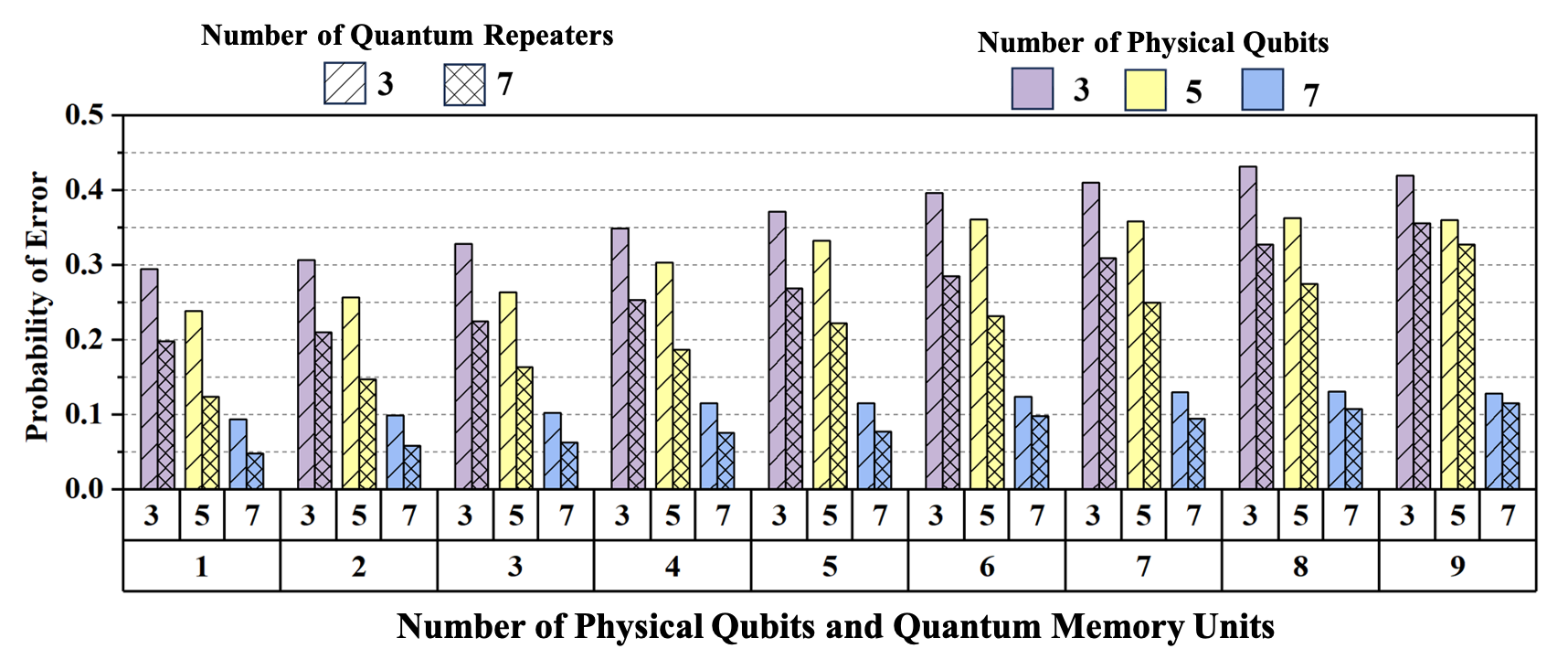}
\label{fig:error_2}}
\caption{\textcolor{black}{Probability of error of logical qubit decoding with (a) direct maximum weight matching and (b) lookup table decoding method. We observe: 1) the utilization of a larger number of quantum repeaters and additional physical qubits results in a reduction of errors; 2) the presence of quantum memory units contributes to increased errors, attributed to the additional waiting time in queuing; 3) the probability of errors shows a strong correlation with the decoding method, for e.g., the lookup-table decoding method is better suited to decrease the probability of errors.}}
\vspace*{-10pt}
\end{figure*}

\begin{figure*}[h!]
\centering
\subfloat[]{
\includegraphics[width=0.3\linewidth]{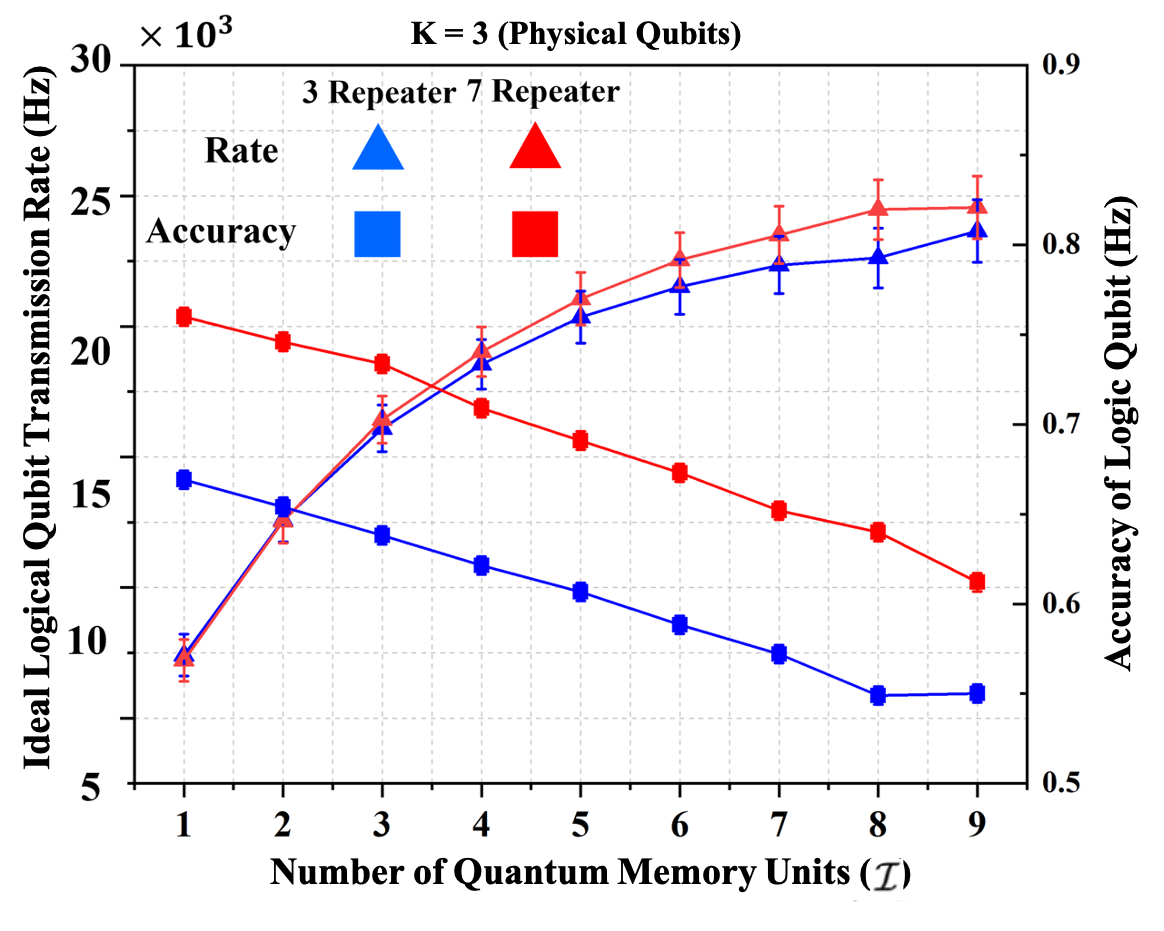}}
\subfloat[]{
\includegraphics[width=0.3\linewidth]{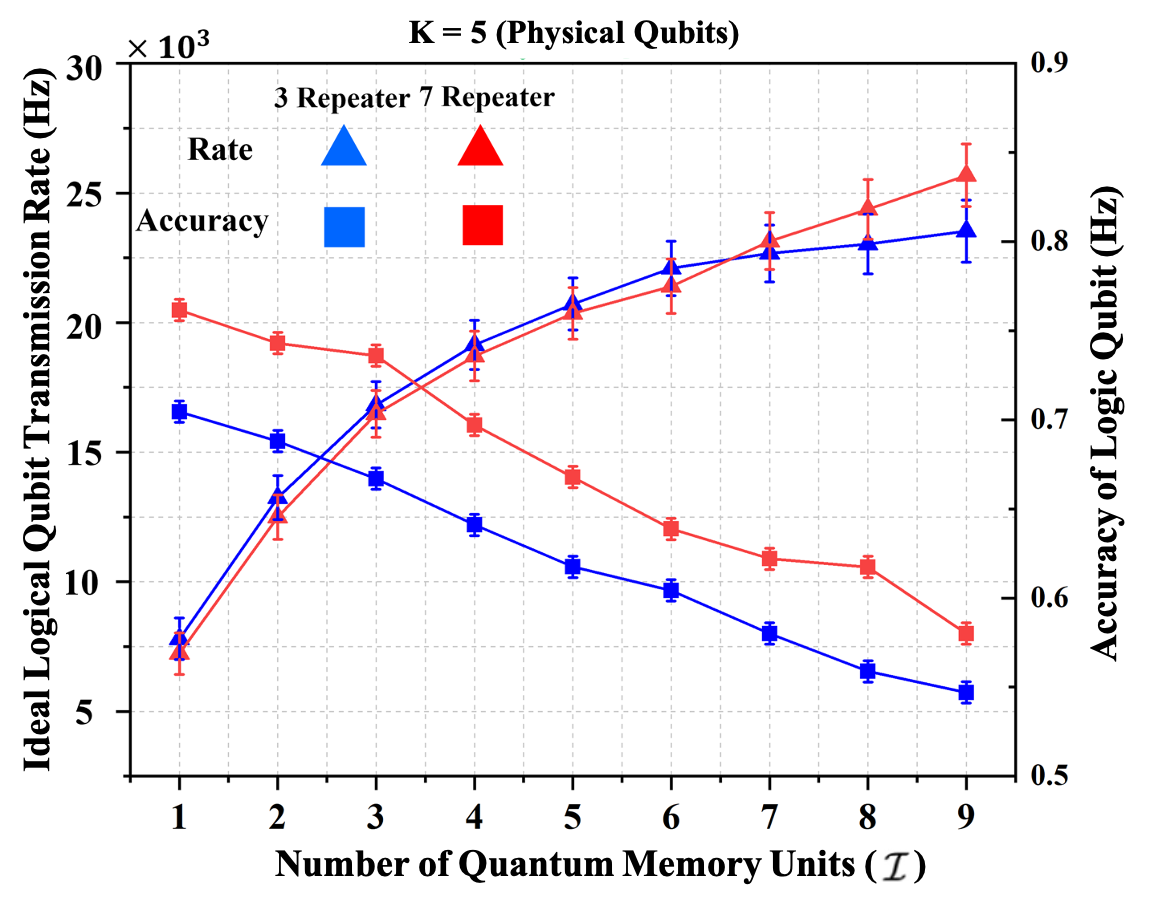}}
\subfloat[]{
\includegraphics[width=0.3\linewidth]{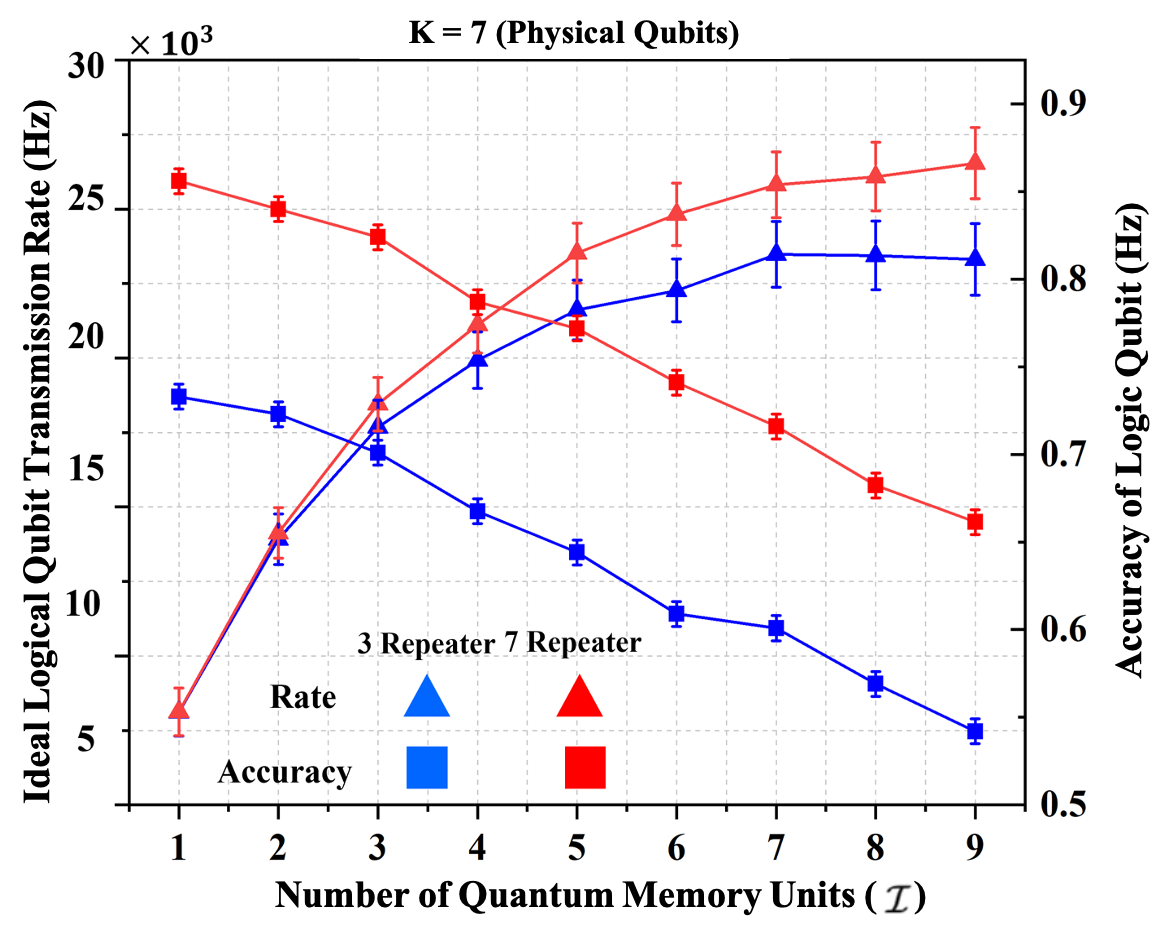}}

\subfloat[]{
\includegraphics[width=0.3\linewidth]{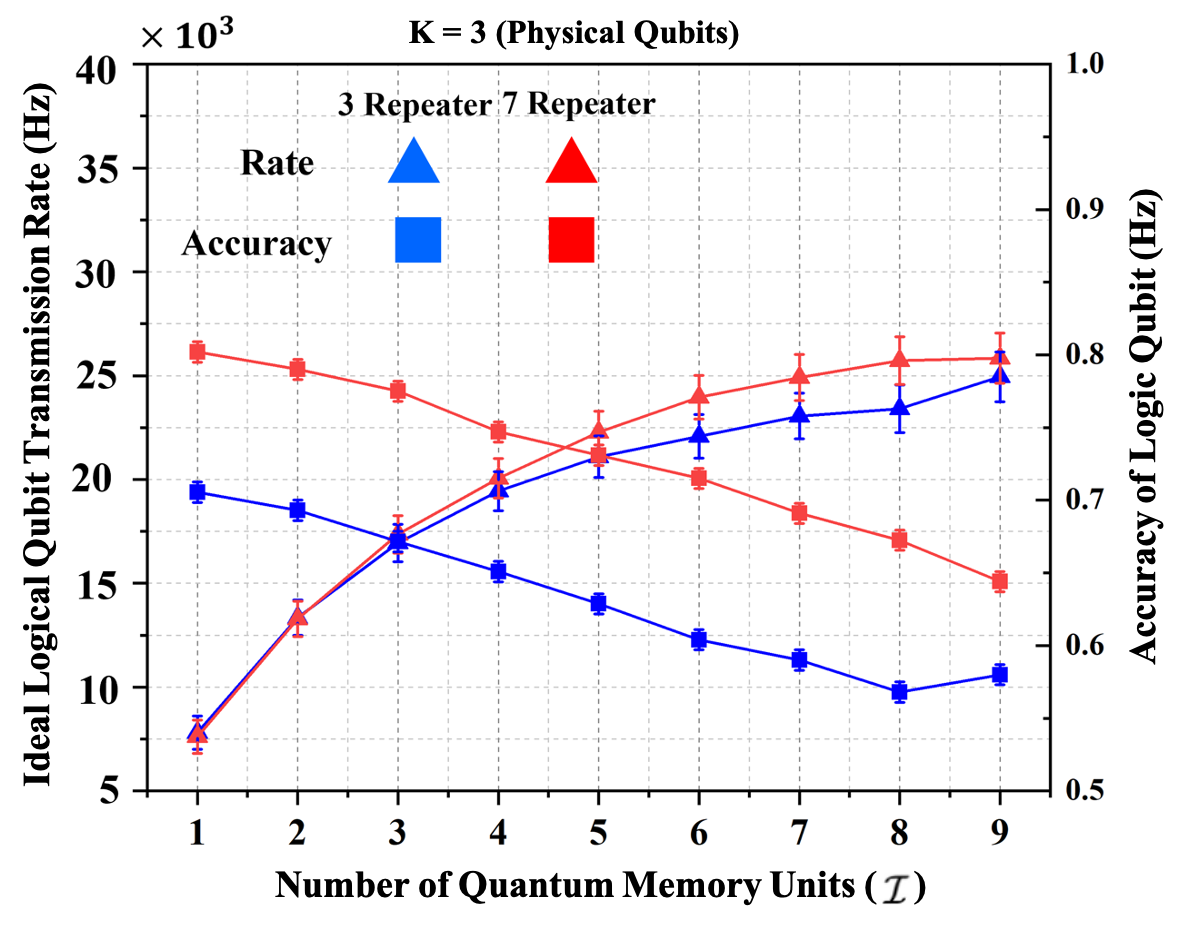}}
\subfloat[]{
\includegraphics[width=0.3\linewidth]{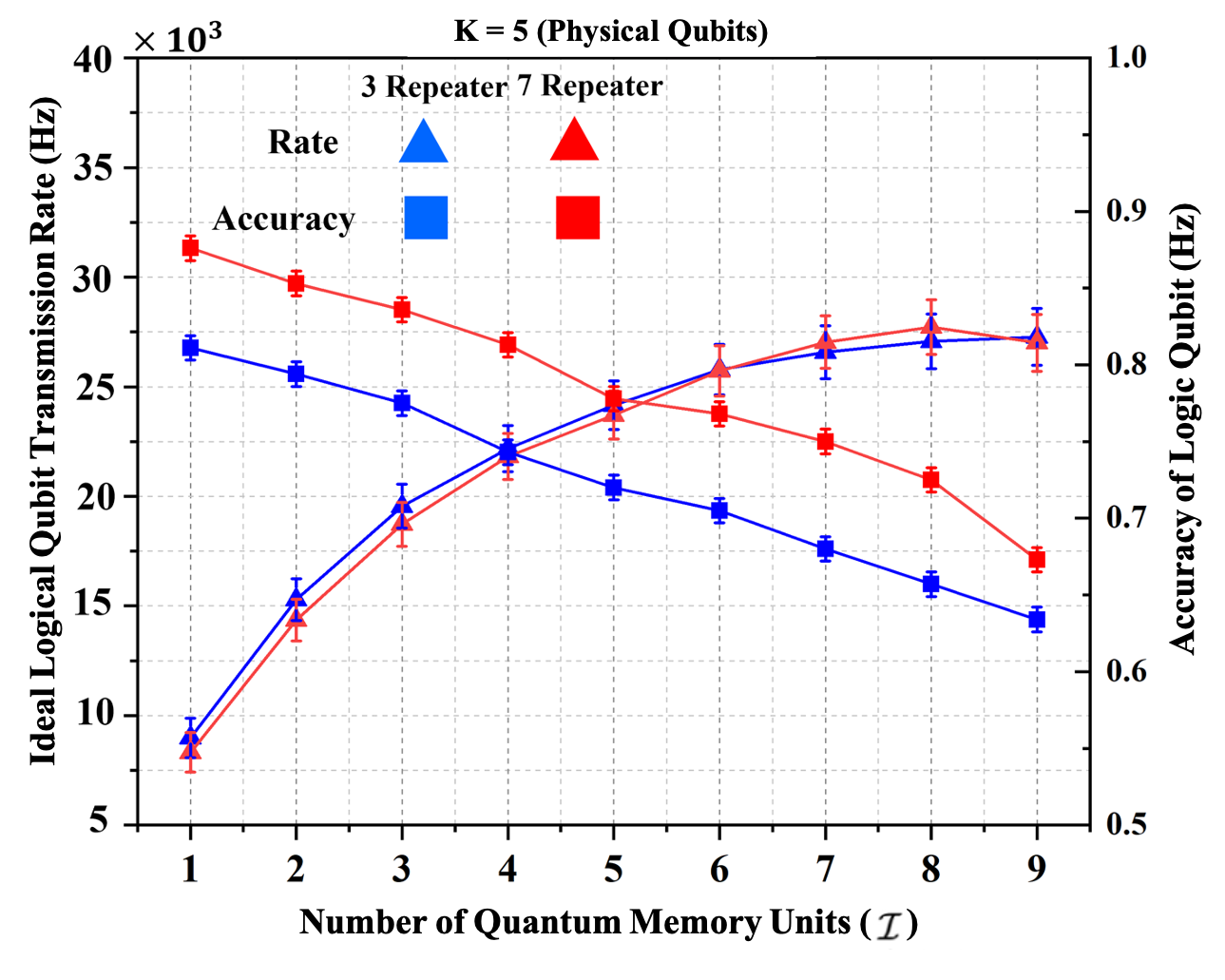}}
\subfloat[]{
\includegraphics[width=0.3\linewidth]{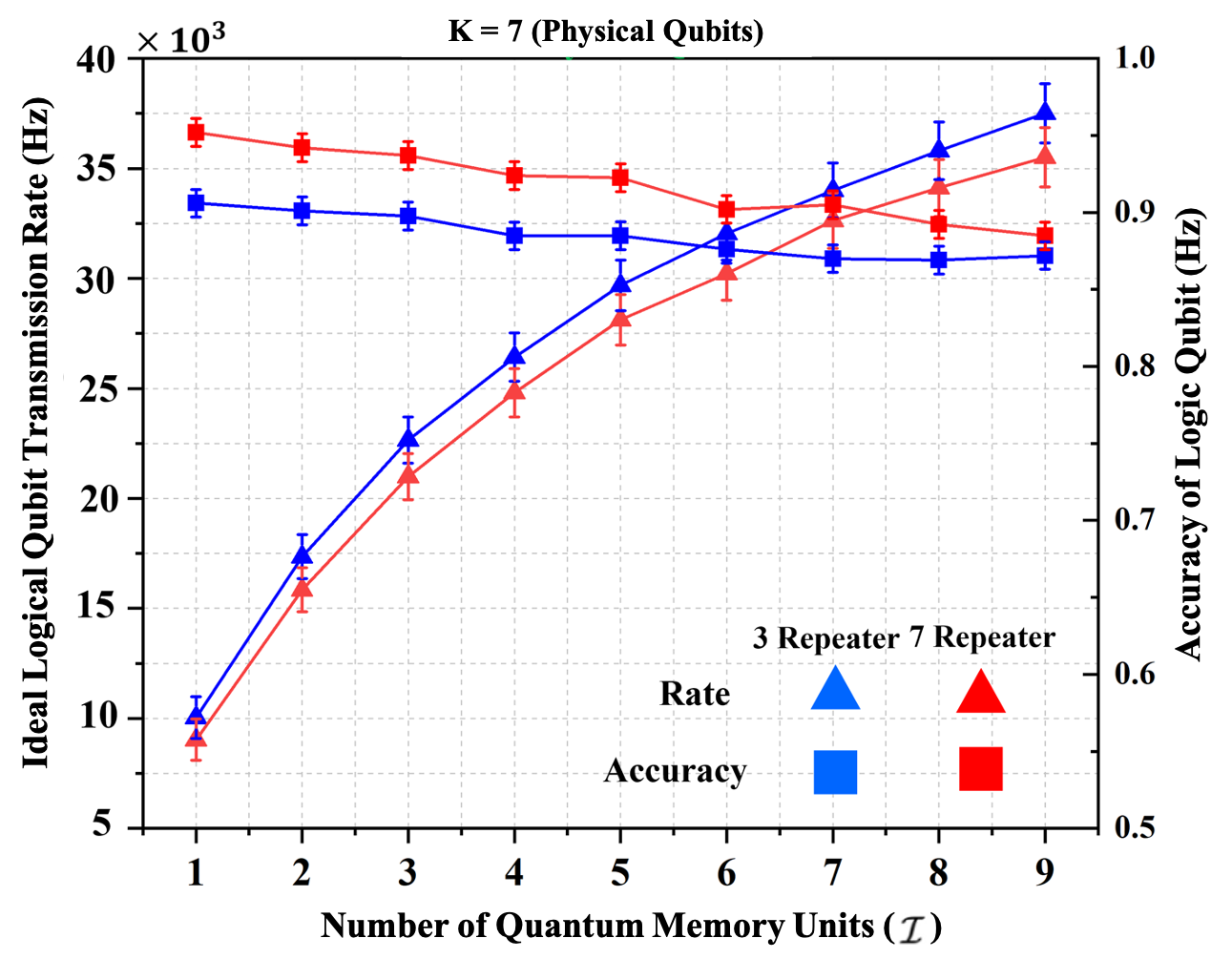}}
\caption{Ideal logical qubit transmission rate with related decoding accuracy: (a)-(c) direct maximum weight matching and (d)-(f) lookup table decoding method. The ideal logical qubit transmission rate exhibits an increasing trend with the addition of quantum memory units and quantum repeaters for two different decoding methods, but the accuracy of the transmitted logical qubit demonstrates an opposite trend. These findings show a trade-off between the rate and accuracy of the transmitted logical qubit. }
\vspace*{-16pt}
\label{fig:rate_accuracy}
\end{figure*}

\subsection{Analysis of Correct Logical Qubit Decoding}
\label{subsec:analysis_correct}
\vspace*{-3pt}
In this section, we analyze the probability of error in the logical qubit. Fig. \ref{fig:q_circuit} shows the quantum circuit used to obtain the syndrome bits through the measurement outcomes of ancillary qubits. These syndrome bits are then utilized to facilitate the decoding of the logical qubits. We utilize the {\tt Qiskit} platform~\cite{qiskit} to design and simulate the circuits. To evaluate the decoding accuracy, we consider a path with a length of $80$ Km, connected by either $3$ or $7$ repeaters. The choice of the number of repeaters ensures that the fidelity of each path remains above a certain threshold (greater than $0.5$) for different numbers of quantum memory units. Furthermore, we consider the logical qubits are encoded using $3$, $5$, and $7$ physical qubits, respectively. This encoding scheme allows us to study the impact of different levels of redundancy on the decoding accuracy and overall performance of the end-to-end route. 

Figs.~\ref{fig:error_1} and \ref{fig:error_2} showcase the variation in the probability of error for decoding of logical qubits across different scenarios. Both decoding methods exhibit lower probabilities of error in scenarios where more quantum repeaters are involved. 
Furthermore, the probability of error increases with the number of quantum memory units, before  stabilizing. This behavior can be attributed to the fidelity (illustrated in Fig. \ref{fig:fidelity}) of each path, which remains close to $0.5$. The decoding method employed in this study approaches the lower bound of the fidelity requirement, thereby resulting in stable error probability. We also observe that the lookup table decoding method exhibits a lower probability of error. However, this method comes at the expense of increased computing resources in the lookup table (the corresponding analysis is out of the scope of this work). 

\begin{observation} 
\vspace*{-2pt}
The probability of error of the logical qubit is strongly correlated with the fidelity of the transmitted qubit in each path, which is also affected by different decoding methods (details in Figs.~\ref{fig:error_1} and \ref{fig:error_2}, validates Contributions 2 and 3). 
\vspace*{-5pt}
\end{observation}

\subsection{Logical Qubit Transmission Rate with Error Correction}
\label{subsec:logical_qubit}
\vspace*{-0pt}

Finally, we evaluate the overall performance of the system, taking into account the ideal logical qubit transmission rate and the corresponding logical qubit decoding accuracy under different decoding methods. The results of the ideal logical qubit transmission rates are calculated by the optimization problem formulated in Eq. \eqref{eq:opt_twin_selection}. In Fig. \ref{fig:rate_accuracy} results (a-c) are  obtained by the direct maximum weight matching based decoding while (d-f) represent the results obtained by the lookup table decoding. We see that the latter gives higher accuracy than the method based on the maximum weight matching. Consequentially, the higher decoding accuracy results in appropriately setting the logical qubit transmission rate, as obtained from Eq.~\eqref{equ:corrected_rate}. Additionally, these evaluation results illustrate the trade-off between the number of memory units and the optimally set qubit transmission rate. Increasing the number of memory units used in each repeater can lead to a higher transmission rate, but it comes at the cost of lower logical qubit decoding accuracy. However, encoding logical qubit with more physical qubits can increase accuracy, for e.g., a logical qubit encoded by $7$ qubits results in better decoding compared to others. Furthermore, in scenarios with more repeaters (e.g., $7$), higher decoding accuracy is observed compared to scenarios with fewer repeaters (e.g., $3$). However, it is important to note that the higher accuracy achieved in the scenario with more repeaters does not always sustain a higher logical qubit transmission rate. This is due to the extra time overhead during the entanglement swapping operation in the scenario with $7$ repeaters, denoted as $t_{w-e}$ in Eq.~\eqref{equ:corrected_rate}. Thus, this time delay can offset the gain in accuracy. % These results encompass a comprehensive analysis of various parameters, yielding diverse system performance outcomes. 
For each scenario, the optimized values can be readily obtained through the formulated optimization problem, for e.g., the optimal logical qubit transmission rate is $>35$ kHz with decoding accuracy of $>0.85$ when using $7$ physical qubits, $9$ quantum memory units and lookup table decoding method from Fig.~\ref{fig:rate_accuracy}f. This enables us to fine-tune the system configuration in terms of selecting ideal logical qubit transmission rate, subject to a target decoding accuracy, and overall efficiency.

\begin{observation}
\vspace*{-3pt}
Different parameters of the proposed caching network 
% including logic qubit transmission rate, decoding accuracy, number of repeater, number of quantuem memory, which 
result in a range of system performance outcomes. The optimized performance can be readily obtained through the formulated optimization problem, which 
% This enables us to fine-tune the system's configuration and achieve optimal performance in terms of logic qubit transmission rate, decoding accuracy, number of repeater, number of quantuem memory and overall efficiency.
% Moreover, these results serve as a benchmark for evaluating the overall system performance. This benchmark 
is crucial for guiding the design and optimization of the quantum caching network (details in Fig.\ref{fig:rate_accuracy}, validates Contributions 2 and 3).
% enabling us to make informed decisions and achieve the desired performance in practical implementations \textcolor{red}{will change a bit}. 
\vspace*{-3pt}
\end{observation}

% corrected logic qubit transmission rate is higher while leveraging the lookup table based the logic qubit decoding has higher accuracy. And, the decoding accuracy  

% These two figures convey two key observations. Firstly, the fidelity consistently decreases with longer transmission lengths, irrespective of the number of quantum memories and quantum repeaters employed. This phenomenon primarily arises due to the overhead costs incurred by increased transmission lengths. Secondly, the fidelity demonstrates an increasing trend when employing more quantum repeaters. However, there is a clear indication of a diminishing rate of change in this upward trend of fidelity with each additional quantum repeater. The observed variation in the trend can be attributed to the increase in time overhead during the entanglement swapping process. Thirdly, as the quantum memory increases, the average fidelity exhibits a slight decrease. This decrease can be attributed to the overhead cost arising from longer waiting times in the queue, as explained in Figure \ref{fig:queue_time}. The fidelity analysis provides valuable insights, and the subsequent section will demonstrate the accuracy of the logic qubit decoding.

% Figure \ref{fig:circuitdiagram} shows component blocks of our proposed low-power SoftSense node. 
%

%% file: section/conclusion.tex
\section{Conclusion}
\label{sec:conc}
\vspace*{-3pt}
In this paper, we present $\name$, a queuing theory-based approach that leverages a quantum caching network to enhance the performance of logical qubit transmission and correction within such a network. We first state the necessary background and define the theories for designing such quantum caching network, involving the error modeling, cost metric definition, and queuing modeling.  Next, we formulate the optimization problem base on queuing theory and logical qubit decoding accuracy. Through simulation results, we validate $\name$ and establish a correlation between the variation in logical qubit transmission rate and the corresponding decoding accuracy. This correlation serves as a benchmark for evaluating and constructing a robust quantum communication network. Designing a quantum network with actual implementations of quantum memory and validating the $\name$ on real quantum repeater network is a  potential future direction.

% \section{Acknowledgements}
% \label{sec:ACK}
% \vspace*{-3pt}
% The authors gratefully acknowledge the support from PHOTONIDS, Inc. 

% The simulation results demonstrate the logic qubit transmission rate variation corresponding decoding accuracy, which benchmark the evaluation metric to build a robust quantum communication network.

%% file: QReach_reference.bbl
%qubit allocation for distributed quantum computing

%% file: QReach.bbl
\begin{thebibliography}{99}



\bibitem{qkd}
P.~W.~Shor and J.~Preskill,
“Simple proof of security of the BB84 quantum key distribution protocol,”
\emph{Physical Review Letters}, vol.~85, no.~2, p.~441, 2000.



\bibitem{teleportation}
S.~Pirandola, J.~Eisert, C.~Weedbrook, A.~Furusawa, and S.~L.~Braunstein,
“Advances in quantum teleportation,”
\emph{Nature Photonics}, vol.~9, no.~10, pp.~641--652, 2015.



\bibitem{zhao2022e2e}
Y.~Zhao, G.~Zhao, and C.~Qiao,
“E2E fidelity aware routing and purification for throughput maximization in quantum networks,”
in \emph{IEEE Conference on Computer Communications (INFOCOM)}, 2022, pp.~480--489.



\bibitem{zeng2022multi}
Y.~Zeng, J.~Zhang, J.~Liu, Z.~Liu, and Y.~Yang,
“Multi-entanglement routing design over quantum networks,”
in \emph{IEEE Conference on Computer Communications (INFOCOM)}, 2022, pp.~510--519.


\bibitem{Li2023}
K.~Li, V.~Chaudhary, S.~G.~Sanchez, and K.~Chowdhury,
“Q-FiRM: Fidelity-based Rate Maximizing Routes for Quantum Networks,”
in \emph{IEEE Consumer Communications \& Networking Conference (CCNC)}, 2023. 


\bibitem{qiao_1}
Y.~Lan, Y.~Zhao, L.~Huang, and C.~Qiao,
“Asynchronous Entanglement Provisioning and Routing for Distributed Quantum Computing,”
in \emph{IEEE Conference on Computer Communications (INFOCOM)}, 2023.




\bibitem{yuanyuan}
M.~Yingling, L.~Yu, and Y.~Yuanyuan,
“Qubit allocation for distributed quantum computing,”
in \emph{IEEE Conference on Computer Communications (INFOCOM)}, 2023.



\bibitem{dist_sensing}
X.~Guo \emph{et al.},
“Distributed quantum sensing in a continuous-variable entangled network,”
\emph{Nature Physics}, vol.~16, no.~3, pp.~281--284, 2020.


\bibitem{quantum_clock}
P.~Komar \emph{et al.},
“A quantum network of clocks,”
\emph{Nature Physics}, vol.~10, no.~8, pp.~582--587, 2014.



\bibitem{chaudhary2023learning}
V.~Chaudhary, K.~Li, and K.~Chowdhury,
“Learning-Based Route Selection in Noisy Quantum Communication Networks,”
in \emph{ICC 2023—IEEE International Conference on Communications}, 2023, pp.~4188--4193.


\bibitem{li2023bip}
K.~Li, V.~Chaudhary, and K.~R.~Chowdhury,
“BiP: Bit-Phase-Flip Error Mitigation in Quantum Communications,”
in \emph{ICC 2023—IEEE International Conference on Communications}, 2023, pp.~4182--4187.



\bibitem{waiting_1}
E.~Shchukin, F.~Schmidt, and P.~van~Loock,
“Waiting time in quantum repeaters with probabilistic entanglement swapping,”
\emph{Physical Review A}, vol.~100, no.~3, p.~032322, 2019.


\bibitem{waiting_2}
S.~Brand, T.~Coopmans, and D.~Elkouss,
“Efficient computation of the waiting time and fidelity in quantum repeater chains,”
\emph{IEEE Journal on Selected Areas in Communications}, vol.~38, no.~3, pp.~619--639, 2020.


\bibitem{waiting_qiao}
Y.~Zhao, G.~Zhao, and C.~Qiao,
“E2E fidelity aware routing and purification for throughput maximization in quantum networks,”
in \emph{IEEE Conference on Computer Communications (INFOCOM)}, 2022, pp.~480--489.


\bibitem{li_waiting}
C.~Li, T.~Li, Y.-X.~Liu, and P.~Cappellaro,
“Effective routing design for remote entanglement generation on quantum networks,”
\emph{npj Quantum Information}, vol.~7, no.~1, p.~10, 2021.


\bibitem{dai2020quantum}
W.~Dai, T.~Peng, and M.~Z.~Win,
“Quantum queuing delay,”
\emph{IEEE Journal on Selected Areas in Communications}, vol.~38, no.~3, pp.~605--618, 2020.



\bibitem{li2020universally}
Y.~Li and S.~Ioannidis,
“Universally stable cache networks,”
in \emph{IEEE Conference on Computer Communications (INFOCOM)}, 2020, pp.~546--555.

\bibitem{das2022lilliput}
P.~Das, A.~Locharla, and C.~Jones,
“LILLIPUT: a lightweight low-latency lookup-table decoder for near-term Quantum error correction,”
in \emph{Proceedings of the 27th ACM International Conference on Architectural Support for Programming Languages and Operating Systems}, 2022, pp.~541--553.


\bibitem{panigrahy2022capacity}
N.~K.~Panigrahy, T.~Vasantam, D.~Towsley, and L.~Tassiulas,
“On the capacity region of a quantum switch with entanglement purification,”
\emph{arXiv preprint} arXiv:2212.01463, 2022.



\bibitem{gera2020mhz}
S.~Gera, S.~Sagona-Stophel, and E.~Figueroa,
“MHz Source of Single Photons Tuned to Rubidium Transition,”
in \emph{Quantum 2.0}, 2020, pp.~QTh7B--17.


\bibitem{lei2022electromagnetically}
X.~Lei, L.~Ma, J.~Yan, X.~Zhou, Z.~Yan, and X.~Jia,
“Electromagnetically induced transparency quantum memory for non-classical states of light,”
\emph{Advances in Physics: X}, vol.~7, no.~1, p.~2060133, 2022.



\bibitem{pouryousef2022quantum}
S.~Pouryousef, N.~K.~Panigrahy, and D.~Towsley,
“A quantum overlay network for efficient entanglement distribution,”
\emph{arXiv preprint} arXiv:2212.01694, 2022.



\bibitem{farahbakhsh2022opportunistic}
A.~Farahbakhsh and C.~Feng,
“Opportunistic routing in quantum networks,”
in \emph{IEEE Conference on Computer Communications (INFOCOM)}, 2022, pp.~490--499.




\bibitem{gottesman1997stabilizer}
D.~Gottesman,
\emph{Stabilizer codes and quantum error correction}.
Ph.D. dissertation, California Institute of Technology, 1997.



\bibitem{shor1995scheme}
P.~W.~Shor,
“Scheme for reducing decoherence in quantum computer memory,”
\emph{Physical Review A}, vol.~52, no.~4, p.~R2493, 1995.


\bibitem{luo2021quantum}
Y.-H.~Luo \emph{et al.},
“Quantum teleportation of physical qubits into logical code spaces,”
\emph{Proceedings of the National Academy of Sciences}, vol.~118, no.~36, p.~e2026250118, 2021.



\bibitem{vardoyan2019stochastic}
G.~Vardoyan, S.~Guha, P.~Nain, and D.~Towsley,
“On the stochastic analysis of a quantum entanglement switch,”
\emph{ACM SIGMETRICS Performance Evaluation Review}, vol.~47, no.~2, pp.~27--29, 2019.



\bibitem{request_schedule}
C.~Cicconetti, M.~Conti, and A.~Passarella,
“Request scheduling in quantum networks,”
\emph{IEEE Transactions on Quantum Engineering}, vol.~2, pp.~2--17, 2021.



\bibitem{lovas2021markov}
A.~Lovas and M.~R{\'a}sonyi,
“Markov chains in random environment with applications in queuing theory and machine learning,”
\emph{Stochastic Processes and their Applications}, vol.~137, pp.~294--326, 2021.



\bibitem{brun2000analytical}
O.~Brun and J.-M.~Garcia,
“Analytical solution of finite capacity M/D/1 queues,”
\emph{Journal of Applied Probability}, vol.~37, no.~4, pp.~1092--1098, 2000.

\bibitem{qiskit}
Qiskit,
“Qiskit,” 2023. [Online]. Available: \url{https://qiskit.org}

\end{thebibliography}
